\documentclass[aps, prb, onecolumn, amssymb, superscriptaddress, nofootinbib]{revtex4-1}
\usepackage{bm, amsmath, amsfonts}
\usepackage{amsthm}
\usepackage{graphicx, color}
\usepackage{tikz}
\usepackage{braket}
\usepackage[colorlinks=true,linktoc=page,citecolor=red,linkcolor=blue]{hyperref}
\def\bra#1{\langle\mbox{$#1$}\rvert}
\def\ket#1{\lvert\mbox{$#1$}\rangle}

\newcommand{\p}{\partial}

\newcommand{\be}{\begin{equation}}      
\newcommand{\ee}{\end{equation}}      
\newcommand{\bea}{\begin{eqnarray}}      
\newcommand{\eea}{\end{eqnarray}}

\newcommand{\im}{\mathrm{i}}
\newcommand{\e}{\mathrm{e}}

\newcommand{\calH}{\mathcal{H}}
\newcommand{\calC}{\mathcal{C}}

\newcommand{\sfR}{\mathsf{R}}

\newcommand{\calV}{\mathcal{V}}

\newcommand{\bZ}{\mathbb{Z}}
\newcommand{\calZ}{\mathcal{Z}}

\newcommand{\bRP}{\mathbb{RP}}
\newcommand{\bCP}{\mathbb{CP}}

\begin{document}

\title{Anomaly indicator of rotation symmetry in (3+1)D topological order}

\author{Ryohei Kobayashi}
\email{r.kobayashi@issp.u-tokyo.ac.jp}
\affiliation{Institute for Solid State Physics, The University of Tokyo, Kashiwa, Chiba 277-8581, Japan}

\author{Ken Shiozaki}
\email{ken.shiozaki@yukawa.kyoto-u.ac.jp}
\affiliation{Yukawa Institute for Theoretical Physics, Kyoto University, Kyoto 606-8502, Japan}

\date{\today}

\begin{abstract}
We examine (3+1)D topological ordered phases with $C_k$ rotation symmetry. 
We show that some rotation symmetric (3+1)D topological orders are anomalous, 
in the sense that they cannot exist in standalone (3+1)D systems, but only exist on the surface of (4+1)D SPT phases.
For (3+1)D discrete gauge theories, we propose anomaly indicator that can diagnose the $\bZ_k$ valued rotation anomaly.
Since (3+1)D topological phases support both point-like and loop-like excitations, 
the indicator is expressed in terms of the symmetry properties of point and loop-like excitations, 
and topological data of (3+1)D discrete gauge theories.
  
\end{abstract}

\maketitle

\tableofcontents

\section{Introduction}
\label{sec:intro}
Understanding possible phases of matter in the presence of symmetries is important in condensed matter physics.
For (2+1)D topological ordered phases with global symmetries~\cite{Wenbook},
the symmetry properties are classified through studying the symmetry action on quasiparticle excitations.
Such symmetry properties on quasiparticles collectively describe symmetry enriched topological (SET) phase.~\cite{wen2002quantum, levin2012classification, essin2013classifying}
Concretely, when the global symmetry $G$ is unitary and onsite, the (2+1)D SET phase is characterized by 
algebraic properties of symmetry defects. Roughly speaking, these properties depict the fusion and braiding data of quasiparticles along with symmetry defects,
which gives rise to enlarged theory of quasiparticles incorporated with $G$-defects. 
Formulation of the enlarged theory by $G$-defects is given by mathematical object known as unitary $G$-crossed braided fusion category~\cite{etingof, barkeshli2014}.
In particular, $G$-crossed braided fusion category encodes the data of 
symmetry fractionalization on quasiparticles, reminiscent of anyons with fractional electric charge in fractional quantum Hall effect 
and $S=1/2$ spinon excitations in quantum spin liquids.

Interestingly, some fractionalization patterns of SET phases in (2+1)D lead to an anomaly, in the sense that they cannot exist as standalone (2+1)D systems, 
but only exist on the surface of (3+1)D symmetry protected topological (SPT) phases~\cite{chen2015anomalous, hermele2016flux, kapustin2014anomalies, benini20182group}.
 The anomaly arises when there is an obstruction to gauging symmetries,
and their presence depends on symmetry fractionalization pattern in the system.
For unitary onsite symmetries, the anomaly manifests itself as obstructions to $G$-crossed extension of braided fusion categories for a given pattern of symmetry fractionalization.
For spacetime symmetries, there has also been a number of works 
examining symmetry fractionalization and anomalies involving time reversal or space group symmetries~\cite{lake2016anomalies, qi2017folding, metlitski2013bosonic, wang2013gapped, metlitski2015symmetry, fidkowski2013non, seibergwitten2016, wang2017anomaly, lee2018study, cheng2017exactly, qi2015anomalous}.
In the case of reflection and time reversal symmetry, conceptual understanding of anomalies in (2+1)D bosonic SET phases is developed in Ref.~\cite{barkeshli2016}.
The (3+1)D bosonic SPT phases protected by reflection or time reversal symmetry are classified as $\bZ_2\times\bZ_2$, 
which corresponds to (3+1)D unoriented bordism group $\Omega_4^O(pt)=\bZ_2\times \bZ_2$~\cite{kapustin2014beyond}.
This bordism group is generated by two manifolds, $\mathbb{RP}^4$ and $\mathbb{CP}^2$. Thus, anomaly on (2+1)D SET phase is detected by 
the partition function of bulk SPT phase on generator manifolds; $\calZ(\mathbb{RP}^4)=\pm1$, $\calZ(\mathbb{CP}^2)=\pm1$. 
In Ref.~\cite{barkeshli2016}, the authors evaluated path integrals of (3+1)D SPT phases on generator manifolds $\bRP^4, \bCP^2$,
based on a given input of topological order on the surface, and computed the SPT partition functions as
\begin{align}
\calZ(\mathbb{RP}^4)&=\frac{1}{\mathcal{D}}\sum_{\bar{p}=\sfR(p)}d_p\eta_pe^{\im\theta_p},
\label{RP4part}
\end{align}
\begin{align}
\calZ(\mathbb{CP}^2)&=\frac{1}{\mathcal{D}}\sum_{p}d_p^2e^{\im\theta_p}=e^{\frac{2\pi\im}{8}c_{-}},
\label{CP2part}
\end{align}
where $d_p$ is quantum dimension of $p$, $\mathcal{D}$ is total dimension characterized by $\mathcal{D}^2:=\sum_pd_p^2$, and $\theta_p$ is $\mathbb{R}/2\pi\bZ$-valued topological spin of $p$.
$\eta_p$ is a $\bZ_2$ valued quantity that characterizes the symmetry fractionalization of a quasiparticle $p$, which will be defined shortly.
In~\eqref{CP2part}, we see that $\calZ(\mathbb{CP}^2)$ is related to the chiral central charge $c_{-}$ of the surface theory.

$\eta_p$ is defined as the $\sfR$ eigenvalue of the reflection symmetric state $\ket{p, \sfR(p)}$, 
where $\sfR$ denotes the reflection, and two quasiparticles $p$, $\sfR(p)$ are located in reflection symmetric fashion. Namely, we have
\begin{align}
\sfR\ket{p, \sfR(p)}=\ket{\sfR^2(p), \sfR(p)}=:\eta_p\cdot\ket{p, \sfR(p)}.
\label{defeta}
\end{align}
In the first equation in~\eqref{defeta}, we note that the reflection permutes the position of two quasiparticles.
$\eta_p$ takes value in $\pm1$ since we have $\sfR^2$=1 on the Hilbert space. The state $\ket{p, \sfR(p)}$ exists only when 
$p, \sfR(p)$ fuse into vacuum; $\overline{p}=\sfR(p)$, otherwise $\eta_p$ becomes ill-defined.
Accordingly, summation runs over quasiparticles such that $\bar{p}=\sfR(p)$ in~\eqref{RP4part}.

These formulae~\eqref{RP4part},~\eqref{CP2part} are sometimes called the ``anomaly indicator"~\cite{wang2017anomaly} 
that allows us to diagnose anomalies from input data of (2+1)D SET phases.
For instance, let us take a look at the toric code with $\sfR$ symmetry. 
Since $\eta$ should be compatible with the fusion rule, we must have $\eta_a\eta_b=\eta_c$ whenever $N_{ab}^{c}\neq 0$.
Hence, the fractionalization is determined by $\eta_e$ and $\eta_m$, when $\sfR$ does not permute anyons.
In this situation, there are four choices of $\eta_p$ summarized in Table~\ref{etatable}.
\begin{table}[htb]
\centering
  \begin{tabular}{|c| |c|c|c|c|} \hline
    SET & $\eta_1$ & $\eta_e$ & $\eta_m$ & $\eta_{\psi}$ \\ \hline \hline
    $e1m1$ &1 &$1$ & $1$ & $1$ \\
    $e1mM$ & 1 & $1$ & $-1$ & $-1$ \\
    $eMm1$ & 1 & $-1$& $1$ & $-1$ \\
    $eMmM$ &1 &$-1$ &$-1$ & $1$ \\ \hline
  \end{tabular}
  \caption{Pattern of the fractionalization of reflection symmetry on quasiparticles of the toric code.
 The case of $\eta_e=\eta_m=-1$ ($eMmM$) becomes anomalous, which is realized on the surface of (3+1)D SPT.}
 \label{etatable}
\end{table}
According to the indicator formula~\eqref{RP4part}, we immediately see that $\calZ(\bRP^4)=-1$ 
for $eMmM$ fractionalization $\eta_e=\eta_m=-1$, otherwise $\calZ(\bRP^4)=1$, from which we conclude that only the $eMmM$ state is anomalous.
The indicator formula~\eqref{RP4part} is also generalized for $\bZ_{16}$ valued anomaly of 
fermionic topological phases with reflection symmetry such that $\sfR^2=1$ (known as class DIII in literature)~\cite{tachikawa2017time, tachikawa2017more}.

In this paper, we examine anomalies of spatial symmetry in (3+1)D bosonic topological ordered phases.
While the (2+1)D topological phases possess only point-like excitations (anyons), the (3+1)D topological phases in general support both point and loop-like excitations~\cite{lan2018classification}.
For instance, for $\bZ_N$ gauge theories we find point-like electric particles $e_n$, and loop-like vortex line excitations $q_m$ labeled by $n,m\in\bZ_N$.

\begin{figure}[htb]
\centering
\includegraphics[bb=0 0 497 207]{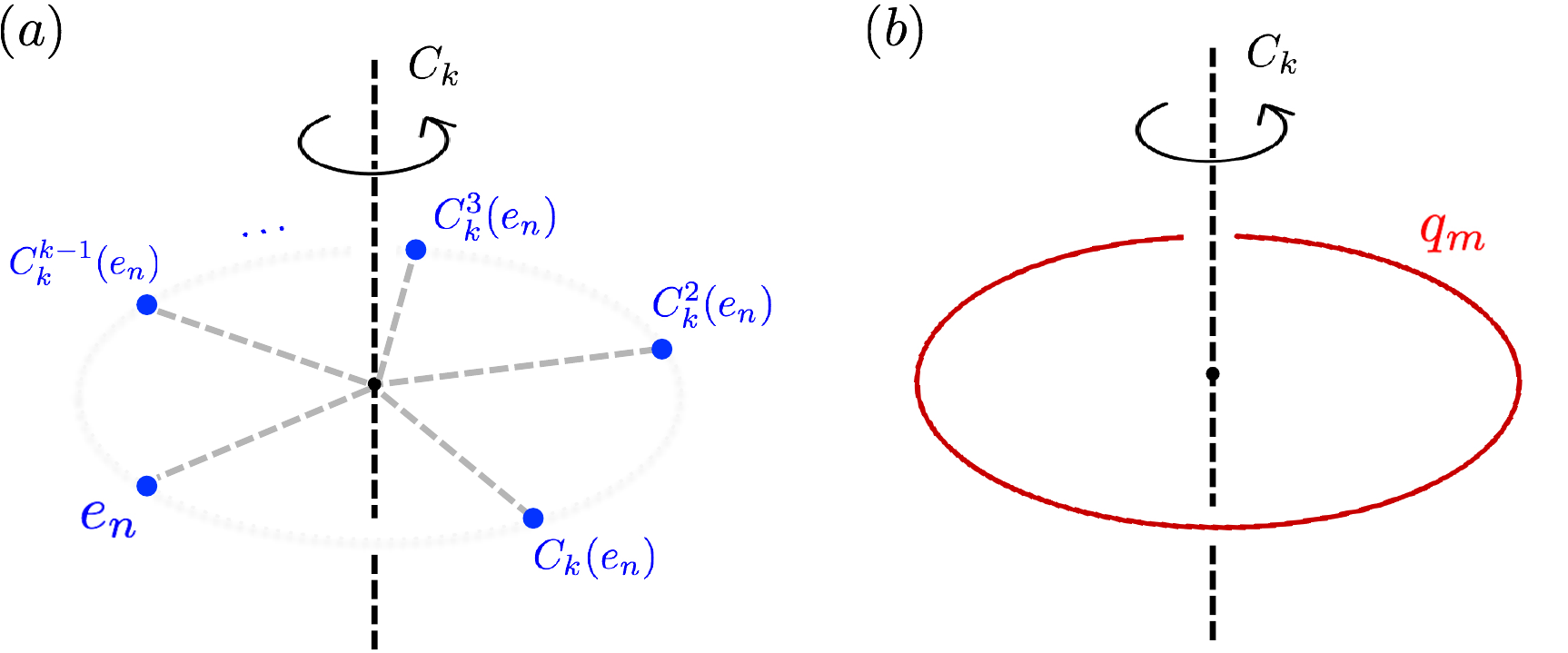}
\caption{$(a)$: $C_k$ symmetric configuration of $k$ point-like excitations. To ensure the existence of such state, 
we must require that $k$ point-like excitations $e_n, C_k(e_n), \dots, C_k^{k-1}(e_n)$ fuse into vacuum.
$(b)$: $C_k$ symmetric configuration of a single loop-like excitation. To ensure the invariance of such state under $C_k$ symmetry, we must require that $C_k(q_m)=q_m$.}
\label{fig:ckaction}
\end{figure}

It is natural to expect that, in (3+1)D topological phases the anomaly is captured by symmetry properties of point-like and loop-like excitations,
as discussed for onsite symmetries in Ref.~\cite{ye2018three, ning2018topological}.
Concretely, we consider the anomaly of $C_k$ rotation symmetry around some axis in (3+1)D untwisted discrete gauge theories 
(i.e., (3+1)D Dijkgraaf-Witten type gauge theory~\cite{dijkgraaf1990topological} with a trivial 4-cocycle).
The anomalies of point group symmetries are systematically classified via the dimensional reduction approach~\cite{song2017topological},
which tells that the $C_k$ anomaly in (3+1)D bosonic systems contains the anomaly that takes the value in $\bZ_k$.
In the case of $\bZ_N$ gauge theories, we propose the indicator formula that can diagnose the $\bZ_k$ part of the total $C_k$ anomaly in (3+1)D as
\begin{align}
\exp\left(\frac{2\pi\im\nu}{k}\right)=\frac{1}{N}\sum_{n,m}\eta_{e_n}\tilde{\eta}_{q_m}\exp\left[-\frac{2\pi\im}{N}n m\right],
\label{abelindicator}
\end{align}
where $\nu\in\bZ_k$ detects the anomaly. Here, $\eta_{e_n}, \tilde{\eta}_{q_m}$ characterize the symmetry fractionalization of the point-like excitation  $e_n$ and 
loop-like excitation (vortex line) $q_m$, respectively. Similar to the case of (2+1)D with $\sfR$ symmetry, $\eta_{e_n}, \tilde{\eta}_{q_m}$ are defined via locating excitations 
in symmetric fashion: $\eta_{e_n}$ is the $C_k$ eigenvalue of the state with $k$ point-like excitations 
$e_n, C_k(e_n), \dots, C_k^{k-1}(e_n)$ located in $C_k$ symmetric fashion (see Fig.\ref{fig:ckaction}.$(a)$).
$\tilde{\eta}_{q_m}$ is the $C_k$ eigenvalue of the state with a single loop-like excitation $q_m$ rounding the rotation axis, located in $C_k$ symmetric fashion (see Fig.\ref{fig:ckaction}.$(b)$).
The sum in~\eqref{abelindicator} runs over point-like excitations $e_n$ such that $k$ particles $e_n, C_k(e_n), \dots, C_k^{k-1}(e_n)$ fuse into vacuum, and loop-like excitations 
$q_m$ such that $q_m=C_k(q_m)$. We can read from the indicator formula that some $C_k$ symmetry action on excitations are prohibited on a standalone (3+1)D system, 
i.e., realized only on the surface of a (4+1)D SPT phase.

As in the case of reflection symmetry in (2+1)D~\cite{barkeshli2016}, we can obatin the indicator formula~\eqref{abelindicator} by evaluating path integral of a (4+1)D SPT phase in the bulk.
In Ref.~\cite{shiozaki2017point, tiwari2017bosonic}, 
the authors construct topological invariants of $(2+1)$D $C_k$ SPT phases by the vacuum expectation value of the ``partial $C_k$ rotation'' operator.
The expectation value is thought of as simulating the path integral of the (2+1)D SPT phase protected by \textit{onsite} $\bZ_k$ symmetry, on the $3$D lens space with flat background $\bZ_k$ gauge field. 
This gives the partition function of (2+1)D $\bZ_k$ SPT phase on the generator manifold. 
In the case of (4+1)D, the classification of SPT phases protected by onsite $\bZ_k$ symmetry contains $\calH^5(\bZ_k, U(1))=\bZ_k$, and such SPT phases are 
detected by the partition function on the 5D lens space $L(k;1,1,1)$ with flat $\bZ_k$ background gauge field.
Thus, we expect that the partition function $\calZ(L(k;1,1,1))$ based on the (4+1)D $C_k$ SPT phase constructed via the partial $C_k$ rotation, 
provides the anomaly indicator of (3+1)D surface topological phases.
After justifying the prediction that $\calZ(L(k;1,1,1))$ gives the topological invariant which distinguishes $C_k$ SPT phases,
we derive the indicator formula~\eqref{abelindicator} by evaluating $\calZ(L(k;1,1,1))$
from a given data of (3+1)D $\bZ_N$ gauge theories on the surface,
\begin{align}
\calZ(L(k;1,1,1))=\frac{1}{N}\sum_{n,m}\eta_{e_n}\tilde{\eta}_{q_m}\exp\left[-\frac{2\pi\im}{N}n m\right].
\end{align}
In addition, generalizing the formula~\eqref{abelindicator}, we pose a conjecture of the $C_k$ anomaly indicator for untwisted non-abelian discrete gauge theories.
The excitations in the discrete $G$-gauge theory are characterized by a vortex line (conjugacy class $\chi$ of $G$) 
and an electric charge (irreducible representation of the centralizer $G_{\chi}$ with respect to $\chi$) attached to the vortex line.
Namely, excitations are labeled by a pair $[\chi, \text{Rep}_i(G_{\chi})]$. In particular, when $\chi=\{1\}$ the excitation $[\chi, \text{Rep}_i(G_{\chi})]$
represents a point-like electric particle labeled by an irreducible representation of $G_{\chi}=G$.
Based on a heuristic argument for evaluating the bulk partition function, we conjecture that the anomaly indicator becomes
\begin{align}
\calZ(L(k;1,1,1))=\frac{1}{\lvert G\rvert}\sum_{\chi, i}\lvert\chi\rvert\dim[\text{Rep}_i(G_{\chi})]\cdot\eta_{\chi; i}\Theta_{\chi; i},
\label{nonabelindicator}
\end{align}
where $\eta_{\chi; i}$ is the $C_k$ eigenvalue of the state with a vortex line $\chi$ and 
$k$ charges $\text{Rep}_i(G_{\chi}), C_k[\text{Rep}_i(G_{\chi})], \dots, C_k^{k-1}[\text{Rep}_i(G_{\chi})]$ located in a $C_k$ symmetric manner. 
$\Theta_{\chi;i}$ denotes the braiding phase between a charge and a vortex line, $\Theta_{\chi;i}:=\text{Rep}_i(G_{h})[h]$, 
where $h\in\chi$ and $G_{h}$ is the centralizer of $h$. 
The sum in~\eqref{nonabelindicator} runs over vortex lines $\chi$ fixed by $C_k$ action $\chi=C_k[\chi]$, and electric charges such that $k$ particles 
$\text{Rep}_i(G_{\chi}), C_k[\text{Rep}_i(G_{\chi})], \dots, C_k^{k-1}[\text{Rep}_i(G_{\chi})]$ fuse into vacuum (i.e., tensor product of $k$ representations contains the trivial representation of $G_{\chi}$). 

The outline of the rest of the paper is summarized as follows. In Section~\ref{sec:dimred}, we first provide the classification of (4+1)D bosonic SPT phases protected by $C_k$ symmetry,
based on the dimensional reduction approach. 
Next, we offer a lattice model of (3+1)D toric code enriched by $C_2$ symmetry, with anomalous symmetry fractionalization under $C_2$ symmetry. 
The constructed model gives the simplest example of surface topological order of (4+1)D $C_2$ SPT phases.
In Section~\ref{sec:indicator}, we first verify that the partition function on the 5D lens space $\calZ(L(k;1,1,1))$ detects 
the (partial) $\bZ_k$ classification of $C_k$ SPT phases in (4+1)D (Sec.~\ref{subsec:partial}), by using the dimensional reduction. 
Next, we derive the indicator formula~\eqref{abelindicator} by explicit computation of $\calZ(L(k;1,1,1))$, for a given $\bZ_N$ gauge theory on the (3+1)D surface.
The computation of $\calZ(L(k;1,1,1))$ is performed by applying gluing relation to the 5D path integral (Sec.~\ref{subsec:gluing}),
which is in parallel with the evaluation of $\calZ(\bRP^4)$ in Ref.~\cite{barkeshli2016}. In Section~\ref{sec:nonabelian}, we pose a conjecture of the anomaly indicator formula 
in (4+1)D untwisted non-abelian discrete gauge theories.

\section{Dimensional reduction approach and anomalous (3+1)D toric code}
\label{sec:dimred}
In this paper, we limit our attention to the anomaly of (3+1)D discrete gauge theories enriched with $C_k$ rotation symmetry. 
Here, we briefly discuss the classification of $C_k$ SPT phases in the (4+1)D bulk, based on the dimensional reduction approach.

Following the logic in Ref.~\cite{song2017topological}, let us think of a small volume $V$ in the spatial manifold of (4+1)D bulk away from the $C_k$ rotation axis 
(in the 4D space, the $C_k$ rotation axis is realized as a 2D plane). Then, we can act $k$ local unitary operators $U_V, U_{C_k(V)}, \dots, U_{C_k^{k-1}(V)}$ 
related by $C_k$ symmetry with each other, which are supported on $V, C_k(V), \dots, C_k^{k-1}(V)$ respectively.
Acting these operators on the SPT state successively gives a $C_k$ symmetric local unitary circuit.
In (4+1)D, the bosonic topological phases without any symmetry are classified as $\bZ_2$, 
which corresponds to the (4+1)D oriented bordism group $\Omega_5^{SO}(pt)=\bZ_2$~\cite{kapustin2014beyond}.
If we do not have the nontrivial SPT phase corresponding to the generator of $\Omega_5^{SO}(pt)=\bZ_2$ in the bulk, 
we can bring the $C_k$ SPT state \textit{away from the rotation axis} to the trivial product state by operating $C_k$ symmetric unitary circuit.

Hence, apart from the nontrivial element of $\Omega_5^{SO}(pt)=\bZ_2$, the (4+1)D $C_k$ SPT phase reduces to the (2+1)D system supported
on the rotation axis, where the $C_k$ symmetry behaves as an onsite $\bZ_k$ symmetry. 
On the (2+1)D rotation axis, there can be a single (2+1)D SPT phase protected by the onsite $\bZ_k$ symmetry (classified as $\calH^3(\bZ_k, U(1))=\bZ_k$), 
or integer $n_{E_8}$ copies of the $E_8$ state~\cite{Kitaevvideo, plamadeala2013short} where the $\bZ_k$ symmetry acts as an identity operator. 
From these states, we obtain $\bZ_k\times\bZ$ classification of the (2+1)D system on the rotation axis.
However, it should be noted that the number of $E_8$ states on the rotation axis $n_{E_8}$ can be changed by $\pm k$, 
by adjoining or annihilating $k$ $E_8$ states with the same chirality in the $C_k$ symmetric way. 
Thus, the classification of the $E_8$ part reduces from $\bZ$ to $\bZ/k\bZ=\bZ_k$.
Eventually, we obtain $\bZ_k\times\bZ_k$ classification on the rotation axis.

The above discussion also gives the classification of $C_k$ anomaly in the (3+1)D surface.
In this paper, we focus on the $\bZ_k$ SPT ($\calH^3(\bZ_k, U(1))=\bZ_k$) part of the total $\bZ_k\times\bZ_k$ anomaly on the $C_k$ rotation axis.
Namely, we consider the $C_k$ anomaly in (3+1)D bosonic systems that is equivalent to the assignment of the (1+1)D boundary of the 
(2+1)D $\bZ_k$ SPT phase located on the $C_k$ rotation axis.

\subsection{anomalous toric code on the (3+1)D surface}
Now, we examine the (3+1)D surface of the (4+1)D $C_k$ SPT phase. Here, the surface can host the topological ordered phase 
enriched with $C_k$ symmetry, where the $C_k$ action is realized on the surface in an anomalous fashion.
As we have seen in $eMmM$ fractionalization of reflection symmetry in (2+1)D toric code (see Table~\ref{etatable}), 
we expect that we can detect anomaly of spatial symmetry in (3+1)D topological phase,
from symmetry fractionalization on point-like and loop-like excitations. 
To see this, we begin with constructing the simplest lattice model of the (3+1)D toric code with anomalous $C_2$ symmetry ($k=2$),
which is a natural generalization of the $eMmM$ toric code in (2+1)D.

Let us consider the (3+1)D toric code on a cubic lattice with anomalous $C_2$ symmetry.
As we have discussed above, the $C_2$ anomaly contains $\calH^3(\bZ_2, U(1))=\bZ_2$ part, 
which is equivalent to the assignment of the (1+1)D boundary of the (2+1)D $\bZ_2$ SPT phase located on the $C_2$ rotation axis.
The construction of anomalous lattice model is based on the effective theory on the boundary of the (2+1)D $\bZ_2$ SPT phase, which is known as the CZX model in literature~\cite{chen2011two}:
we construct the (3+1)D lattice model with anomalous symmetry action, by putting the boundary of the CZX model on the $C_2$ rotation axis.
The similar construction of $eMmM$ toric code in (2+1)D based on dimensional reduction approach is found in Ref.~\cite{song2017topological}.

\begin{figure}[htb]
\centering
\includegraphics[bb=0 0 221 224]{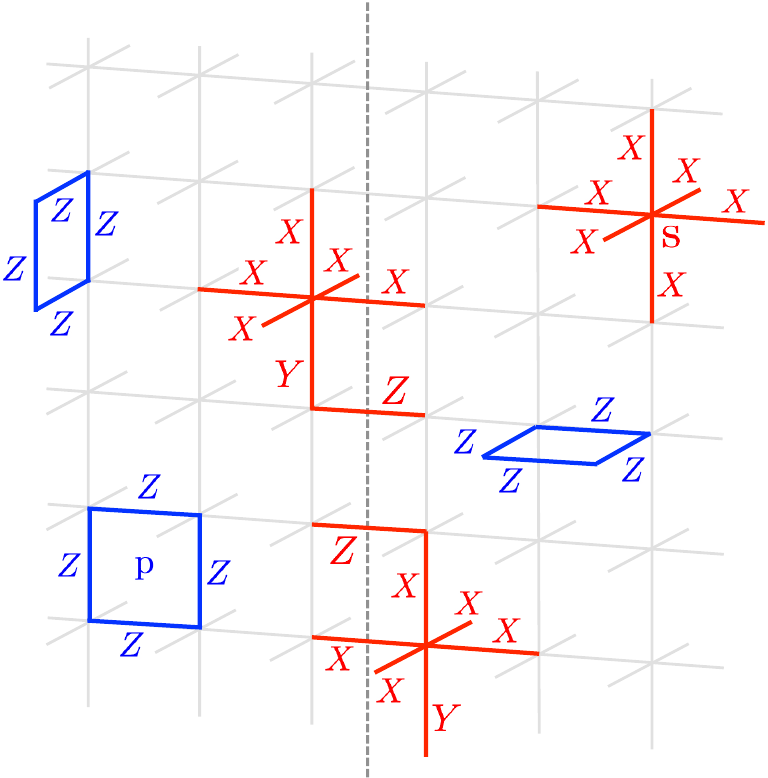}
\caption{3d toric code on a cubic lattice. Qubits on the $C_2$ rotation line (gray dotted line) transform in anomalous way (\ref{toricC2}) under $C_2$ symmetry. 
To respect symmetry, $A_s$ operators (red operators) touching the reflection line are modified. $B_p$ operators (blue operators) need not be modified. }
\label{fig:eCmClat}
\end{figure}

For constructing lattice model, the rotation axis is defined such that the axis intersects edges of cubic lattice, see Fig.\ref{fig:eCmClat}.
The symmetry action is realized on the rotation axis $x=y=0$ as the boundary of CZX model~\cite{chen2011two},
\begin{align}
C_2: X_z\mapsto Z_{z-1}X_{z}Z_{z+1}, \quad Y_{z}\mapsto -Z_{z-1}Y_{z}Z_{z+1}, \quad Z_{z}\mapsto -Z_{z}\qquad \text{at }x=y=0,
\label{toricC2}
\end{align}
otherwise, we have non-anomalous symmetry action. For convenience, we define
\begin{align}
C_2: X_{\vec{x}}\mapsto X_{-\vec{x}}, \quad Y_{\vec{x}}\mapsto -Y_{-\vec{x}},\quad Z_{\vec{x}}\mapsto -Z_{-\vec{x}}, \qquad \text{otherwise.}
\label{toricC22}
\end{align}
Namely, away from rotation axis $C_2$ symmetry acts as the $\pi$ rotation around $x$ axis on qubits.
The anomalous nature of the model is encoded in the non-onsite symmetry action~\eqref{toricC2} realized on the axis.
Now let us define the Hamiltonian which respects the $C_2$ symmetry defined above. 
The Hamiltonian of the conventional toric code has the form of
\begin{align}
H=-\sum_{s}A_s-\sum_{p}B_p,
\label{3dtoriccode}
\end{align}
where $A_s:=\prod_{s\in \partial i}X_i$ is the product of six $X$ operators at edges touching a vertex $s$, 
and $B_p:=\prod_{i\in\partial p}Z_i$ is the product of four $Z$ operators at edges rounding a plaquette (2D square) $p$. 
Although the model (\ref{3dtoriccode}) manifestly does not respect the $C_2$ symmetry~\eqref{toricC2},~\eqref{toricC22},
one can construct the symmetric commuting projector model, just by slightly deforming local Hamiltonians touching the rotation axis.
Let us modify the local Hamiltonian $A_{s}$ adjacent to the rotation axis as
\begin{align}
\begin{cases}
A_s=\left[\prod'_{s\in\partial i}X_i\right]Y_{s-\vec{z}/2}Z_{s+\vec{x}/2-\vec{z}}, &\qquad s_x=-1/2, s_y=0, \\
A_s=-\left[\prod'_{s\in\partial i}X_i\right]Y_{s-\vec{z}/2}Z_{s-\vec{x}/2+\vec{z}}, &\qquad s_x=1/2, s_y=0, 
\end{cases}
\end{align}
where $\left[\prod'_{s\in\partial i}X_i\right]$ stands for product of $X$ at five edges touching a vertex $s$, except for the edge below $s$.
We illustrate the redefined operators in Fig.\ref{fig:eCmClat} (a). 
We can see that the above modification provides 
a commuting projector model which respects the $C_2$ symmetry (\ref{toricC2}),~\eqref{toricC22}.

\begin{figure}[htb]
\centering
\includegraphics[bb=0 0 514 236]{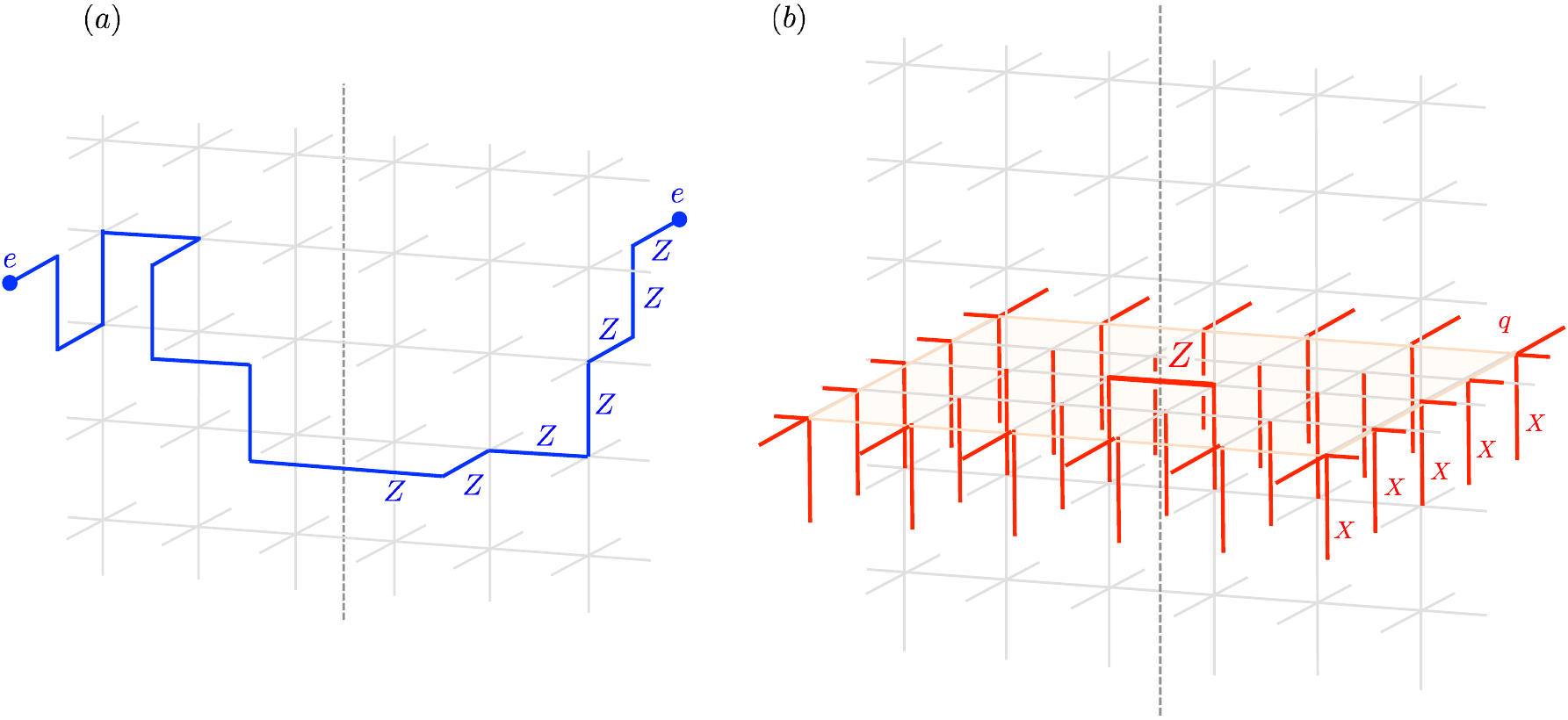}
\caption{$(a)$: $S_e$ string operator which creates a pair of electric $e$ particles at the ends of an open string. 
If two $e$ particles are created in a $C_2$ respecting way, the length of $S_e$ operators are always odd. 
Since we have $C_2: Z\mapsto -Z$ everywhere, we have $S_e\mapsto -S_e$ under $C_2$.
$(b)$: $S_q$ surface operator which creates a loop-like excitation $q$ at the boundary of an open surface. 
$S_q$ operator mostly consists of $X$ operators, but we have a $Z$ operator at the intersection between $S_q$ surface and the $C_2$ axis.
Thus, we have $S_q\mapsto -S_q$ under $C_2$ when $q$ loop rounds the $C_2$ axis.}
\label{fig:em3d}
\end{figure}

Let us examine quasiparticle excitations in the (3+1)D toric code. The (3+1)D toric code has one point-like electric particle
and one loop-like vortex line. Electric particle $e$ violates the Gauss law; $A_s=-1$ if $e$ lives at the vertex $s$. 
Electric particles are generated by acting 1D open line operator $S_e$, which is given by the product of $Z$ operators along the line. 
A pair of particles is created at the ends of an open string. On the other hand, a loop-like vortex line is generated by acting 2D surface operator $S_q$ with a boundary. 
A single vortex line is created at the boundary of an open surface (see Fig.\ref{fig:em3d}).
One can also think of composite excitation of $e$ and  $q$.

How the $C_2$ symmetry acts on $e$ and $q$ particles? As introduced in Sec.~\ref{sec:intro}, 
the symmetry fractionalization $\eta_e, \tilde{\eta}_q$ are defined via locating quasiparticles in a $C_2$ symmetric fashion.
This is performed by acting a line operator $S_e$ and a surface operator $S_q$ as illustrated in Fig.~\ref{fig:em3d}.
$S_e$ creates two $e$ particles in a rotation symmetric way (in this model we have $C_2(e)=e$), 
and $S_q$ creates a single $q$ vortex line rounding the rotation axis.
We can read the symmetry fractionalization $\eta_e$ and $\tilde{\eta}_q$ by the symmetry action on $S_e$ and $S_q$ operators,
$C_2: \ S_e\mapsto \eta_e S_e,\ S_q\mapsto \tilde{\eta}_q S_q$. In the case of our toric code model with anomalous $C_2$ symmetry,
we can show that $S_e$ and $S_m$ acts under $C_2$ as
\begin{align}
C_2:\quad S_e\mapsto -S_e,\quad S_q\mapsto -S_q, 
\end{align}
hence we have $\eta_e=\tilde{\eta}_q=-1$. Among the four choices of symmetry fractionalization $\eta_e=\pm1, \tilde{\eta}_q=\pm1$, 
the three other than $\eta_e=\tilde{\eta}_q=-1$ are easily shown to be realized in a standalone (3+1)D system. 
According to the above observation that $\eta_e=\tilde{\eta}_q=-1$ fractionalization is realized on surface of the (4+1)D $C_2$ SPT phase,
we expect that $\eta_e=\tilde{\eta}_q=-1$ fractionalization pattern is anomalous, otherwise non-anomalous. 
This prediction is immediately confirmed by computing the anomaly indicator~\eqref{abelindicator}, which will be derived in Section~\ref{sec:indicator}.
For instance, for $\eta_e=\tilde{\eta}_q=-1$, the indicator~\eqref{abelindicator} for the $\bZ_2$ gauge theory (toric code, $N=2$) with $C_2$ rotation ($k=2$) becomes,
\begin{align}
\e^{\pi\im\nu}=\frac{1}{2}\sum_{n,m}\eta_{e_n}\tilde{\eta}_{q_m}\e^{-\pi\im nm}=-1,
\end{align}
where $e_0, q_0$ are trivial excitation and $e_1=e$, $q_1=q$ in this expression. Hence, we read from the indicator that the anomaly is $\nu=1$ mod 2.

\section{Anomaly indicator in (3+1)D $\bZ_N$ gauge theory}
\label{sec:indicator}
In this section, we derive anomaly indicator~\eqref{abelindicator} for $G=\bZ_N$ gauge theory in (3+1)D.
After reviewing basic properties of $\bZ_N$ gauge theory in Sec.~\ref{subsec:zngauge}, 
we justify that the partition function on the 5D lens space $\calZ(L(k;1,1,1))$ detects 
the (partial) $\bZ_k$ classification of $C_k$ SPT phases in (4+1)D (Sec.\ref{subsec:partial}), by using the dimensional reduction. 
In Sec.~\ref{subsec:Ck}, we derive the indicator formula~\eqref{abelindicator} by explicit computation of $\calZ(L(k;1,1,1))$, for given $\bZ_N$ gauge theory on the (3+1)D surface.
The computation of $\calZ(L(k;1,1,1))$ is performed by applying gluing relation to the 5D path integral (Sec.~\ref{subsec:gluing}).

\subsection{$\bZ_N$ gauge theories on lattice}
\label{subsec:zngauge}
Discrete gauge theory is formulated on a (3+1)D lattice, whose vertices are labeled by $i$. 
The degrees of freedom live on edges labeled by $ij$. On an oriented edge $ij$, 
such degrees of freedom are discrete $G$-gauge field $g_{ij}\in G$, which satisfies $g_{ij}=g_{ji}^{-1}$.
Vacuum is given by an assignment of a flat $G$-gauge field $\{g_{ij}\}$ on the whole lattice, 
up to gauge equivalence. There are two kinds of extended operators in (3+1)D discrete gauge theory: 
line operator and surface operator (which corresponds to $S_e$ and $S_q$ in the toric code, respectively).

In Section~\ref{sec:nonabelian}, we also pose a conjecture for anomaly indicator for general (possibly non-abelian) gauge theories based on heuristic argument.
In the case of $\bZ_N$ gauge theories, both line and surface operators are labeled by group elements. 
\begin{itemize}
\item The line operator $W_{A,n}(C)$ supported on a closed string on the lattice $C$,
 is labeled by a group element $n\in\bZ_N$:
\begin{align}
W_{A, n}(C)\ket{\{g_{ij}\}}=\exp\left[\frac{2\pi n\im}{N}\prod_{ij\in C}g_{ij}\right]\ket{\{g_{ij}\}},
\end{align}
i.e., a line operator is an Wilson line of electric charge labeled by $n\in\bZ_N$.
If supported on an open string, such line operator generates a pair of point-like electric excitations $e_n, e_{-n}$ at the end of the string.

\item The surface operator $W_{B, m}(S)$ supported on a face of the dual lattice $S$, 
is characterized by a group element $m\in\bZ_N$:
\begin{align}
W_{B, m}(S)=\prod_{ij\in S}\hat{B}_{ij}(m),
\end{align}
where $\hat{B}_{ij}(m)$ is defined as
\begin{align}
\hat{B}_{ij}(m)\ket{g_{ij}}=\ket{m+g_{ij}},
\end{align}
where $i$ is on one side of the surface, and $j$ is on the other side of the surface.
If supported on an open surface, such surface operator generates a loop-like excitation $q_{m}$ at the boundary of the surface.
\end{itemize}

We denote quasiparticles created by $W_{A, n}, W_{B, m}$ as $e_n, q_m$ respectively.
The braiding between $e_n$ and $q_m$ is implemented by the correlator of the line and surface operators 
\begin{align}
\langle W_{A, n}(C) W_{B, m}(S) \rangle=\exp[-\im\frac{2\pi}{N}nm\cdot\text{Lk}(C, S)],
\label{BFbraid}
\end{align}
where $\text{Lk}(C, S)$ denotes the linking number between $C$ and $S$.

\subsubsection{$C_k$ symmetry}
Here, let us briefly refer to properties of $C_k$ symmetry in (3+1)D topological ordered phases.
In general, $C_k$ symmetry can permute the label of quasiparticles. 
For (2+1)D, the symmetry action on anyon labels is defined such that the symmetry leaves the fusion and braiding data invariant, which is formulated as an automorphism of 
unitary braided fusion categories.
\footnote{for orientation reversing symmetry such as $\sfR$, we take symmetry as anti-automorphism instead, which operates on anyon diagrams as 
complex conjugate associated with usual automorphism.}
However, in (3+1)D we generally do not know how to characterize ``automorphism'',
since we do not know what the complete input data is like that can characterize (3+1)D topological ordered phase.
(For instance, diverse link invariants are known in (3+1)D topological ordered phase, see Ref.~\cite{putrov2017braiding}.)

If we limit ourselves to $\bZ_N$ gauge theories, we just have to require that $C_k$ leaves invariant the data of fusion, and the linking phase~\eqref{BFbraid}
between loop and point-like particles.
Concretely, let us assume that $C_k$ acts on the labels as
\begin{align}
C_k: e_1\mapsto e_r, \ q_1\mapsto q_s.
\label{ckaction1}
\end{align}
Since $C_k$ preserves the fusion of quasiparticles, the $C_k$ action on any quasiparticles $e_n, q_m$ are determined by~\eqref{ckaction1}, 
\begin{align}
C_k: e_n\mapsto e_{rn}, \ q_m\mapsto q_{sm}.
\end{align}
The above $C_k$ action induces permutation of quasiparticle labels. Hence, we must have
\begin{align}
\gcd(r, N)=\gcd(s, N)=1.
\label{gcdcond}
\end{align}
In addition, $C_k$ preserves the braiding between point and loop-like excitations~\eqref{BFbraid}. Thus, we must have
\begin{align}
\exp[-\im\frac{2\pi}{N}\cdot\text{Lk}(C, S)]=\exp[-\im\frac{2\pi}{N}rs\cdot\text{Lk}(C, S)],
\end{align}
hence
\begin{align}
rs=1\quad\mathrm{mod}\  N.
\label{braidcond}
\end{align}
Moreover, since $(C_k)^k=1$ on labels, we must have
\begin{align}
r^k=1, \quad s^k=1\quad\mathrm{mod}\  N.
\label{ck1cond}
\end{align}

\subsection{topological invariant of bulk $C_k$ SPT phases via partial rotation}
\label{subsec:partial}
In this subsection, we justify that $\calZ(L(k; 1,1,1))$ works as a topological invariant that diagnoses $\bZ_k$-valued anomaly of $C_k$ symmetry in (3+1)D.
To do this, we first express the partition function $\calZ(L(k; 1,1,1))$ in terms of the ground state expectation value of the partial rotation operator $C_k[D^2\times D^2]$~\cite{shiozaki2017point},
\begin{align}
\calZ(L(k; 1,1,1))=\bra{\Psi_{S^4}}C_{k; 12}[D^2\times D^2]\cdot C_{k; 34}[D^2\times D^2]\ket{\Psi_{S^4}}.
\label{partialrot}
\end{align}
Let us explain the notations in~\eqref{partialrot}. We prepare the Hilbert space on $S^4$, and the $C_k$ SPT ground state on $S^4$ is expressed as $\ket{\Psi_{S^4}}$.
We write the coordinate of $S^4$ as $(x,y,z,w)\in\mathbb{R}^4$, with infinite points identified. $C_k$ transformations are defined as
\begin{align}
C_{k; 12}: \left((x,y),(z,w)\right)\mapsto \left((x,y), C_k(z,w)\right), \quad C_{k; 34}: \left((x,y),(z,w)\right)\mapsto \left(C_k(x,y),(z,w)\right).
 \end{align}
We can think of acting the rotations ``partially'', on $D^2\times D^2: x^2+y^2\le 1, z^2+w^2\le 1$ in $S^4$. 
Let us define the partial rotation operators supported on the $D^2\times D^2$ as $C_{k; 34}[D^2\times D^2], C_{k;12}[D^2\times D^2]$ respectively.
Then, the expectation value of the partial rotations on $D^2\times D^2$~\eqref{partialrot} simulates the path integral on the 5D lens space,
where inserting the partial rotation operator $C_{k; 12}C_{k; 34}$ on a time slice works as creating a ``cross-cap'' in the spacetime. 
Since the lens space $L(k; 1,1,1)$ is defined as identifying two $D^4=D^2\times D^2$ on the boundary of $D^5$ ($S^4=D^4\cup D^4=\p D^5$)
by using the $C_{k; 12}C_{k; 34}$ transformation, inserting the cross-cap makes the spacetime the lens space $L(k; 1,1,1)$. 
(The definition of the 5D lens space is illustrated in Sec.~\ref{subsub:handle}.)

Here, it should be emphasized that one of the $C_k$ transformations $C_{k; 34}, C_{k; 12}$ is the $C_k$ symmetry that is used to define the $C_k$ SPT phase.
The other one is rather taken as an \textit{inherent} $C_k$ symmetry which is the subgroup of $SO(d+1)$ Lorentz symmetry present in TQFT, which is not relevant to symmetry protection.
Hence, we set the $C_{k; 12}$ as a symmetry that protects our SPT phase, and the $C_{k; 34}$ as an inherent one. 

Next, let us perform the dimensional reduction in terms of the $C_{k; 12}$ symmetry. 
As we have explained in Sec.~\ref{sec:dimred}, one can trivialize the $C_{k; 12}$ SPT phase away from the $C_{k; 12}$ rotation axis,
using the dimensional reduction by $C_{k; 12}$ symmetric unitary circuits.
In our case, the $C_{k; 12}$ rotation axis is realized as $S^2: z=w=0$ ($xy$-plane), and we are interested in the $C_{k; 12}$ SPT phase 
that is equivalent to locating a (2+1)D onsite $\bZ_k$ SPT phase on the $xy$-plane.

After the dimensional reduction, the rotation operator $C_{k; 12}$ becomes 
the generator $U_{\bZ_k}$ of the onsite $\bZ_k$ symmetry in the reduced (2+1)D SPT phase.
Thus, the partial rotation operator $C_{k; 12}[D^2\times D^2]$ gives a partial onsite $\bZ_k$ transformation $U_{\bZ_k}[D^2]$ supported on $D^2: x^2+y^2\le1$ in the $xy$-plane,
while $C_{k; 34}[D^2\times D^2]$ still works as a partial rotation operator $C_k[D^2]$ supported on $D^2: x^2+y^2\le1$. 
Therefore, the expectation value in (4+1)D~\eqref{partialrot} reduces to
\begin{align}
\calZ(L(k; 1,1,1))=\bra{\Psi_{S^2}}(U_{\bZ_k}C_k)[D^2]\ket{\Psi_{S^2}},
\label{partialrot2d}
\end{align}
where $\ket{\Psi_{S^2}}$ is the ground state of the (2+1)D $\bZ_k$ SPT phase, whose spatial manifold is taken as $S^2$. 
Now, the expectation value of the partial operation $U_{\bZ_k}C_k$ simulates the partition function of the (2+1)D $\bZ_k$ SPT phase on the 3D lens space $L(k;1,1)$,
in the presence of the flat background $\bZ_k$ gauge field;
\begin{align}
\bra{\Psi_{S^2}}(U_{\bZ_k}C_k)[D^2]\ket{\Psi_{S^2}}=\calZ(L(k;1,1))[A],
\label{z3dlens}
\end{align}
where $A$ denotes the $\bZ_k$ flat background gauge field that corresponds to the generator of $\mathrm{Hom}(\pi_1(L(k;1,1)), \bZ_k)$.
Here, inserting $C_k[D^2]$ creates a cross-cap to make the geometry of the spacetime the lens space $L(k;1,1)$, 
and inserting the $\bZ_k$ symmetry defect $U_{\bZ_k}[D^2]$ on the cross-cap introduces a nontrivial $\bZ_k$ flat connection $A$.
Eventually,~\eqref{z3dlens} gives the partition function of the (2+1)D $\bZ_k$ SPT phase on a generator manifold $\calZ(L(k;1,1))[A]$~\cite{tantivasadakarn2017dimensional}, 
hence detects the distinct $\bZ_k$ SPT phases characterized by $\calH^3(\bZ_k, U(1))=\bZ_k$. 
Therefore, $\calZ(L(k; 1,1,1))$ also diagnoses the $\bZ_k$ classification of (4+1)D $C_k$ SPT phases.

\subsection{gluing relation}
\label{subsec:gluing}
We compute the partition function on a 5D manifold of rather complicated shape (such as the lens space), 
by decomposing the 5D manifold into simpler manifolds which are easier to evaluate, 
and computing the partition function part by part. This procedure is performed via applying the gluing relation for the path integral.
Here, let us review some axiomatic properties of path integral for topological field theories,
which is required for explicit computations, following Ref.~\cite{barkeshli2016, Walker06}.

To consider the path integral on a $(d+1)$D manifold $M^{d+1}$, 
we first specify the configuration of fields on boundary $c\in\calC(\p M^{d+1})$, where $\calC(\p M^{d+1})$ denotes a set of boundary conditions.
If a $d$D manifold $M^d$ has a boundary, we denote $\calC(M^{d}; c)$ as the configuration space of boundary conditions on $M^{d}$, which is fixed as $c$ on $\p M^d$.
In our case where the surface theory is described by discrete gauge theory, a configuration is an assignment of flat $G$-gauge fields on boundaries.

\begin{figure}[htb]
\centering
\includegraphics[bb=0 0 449 215]{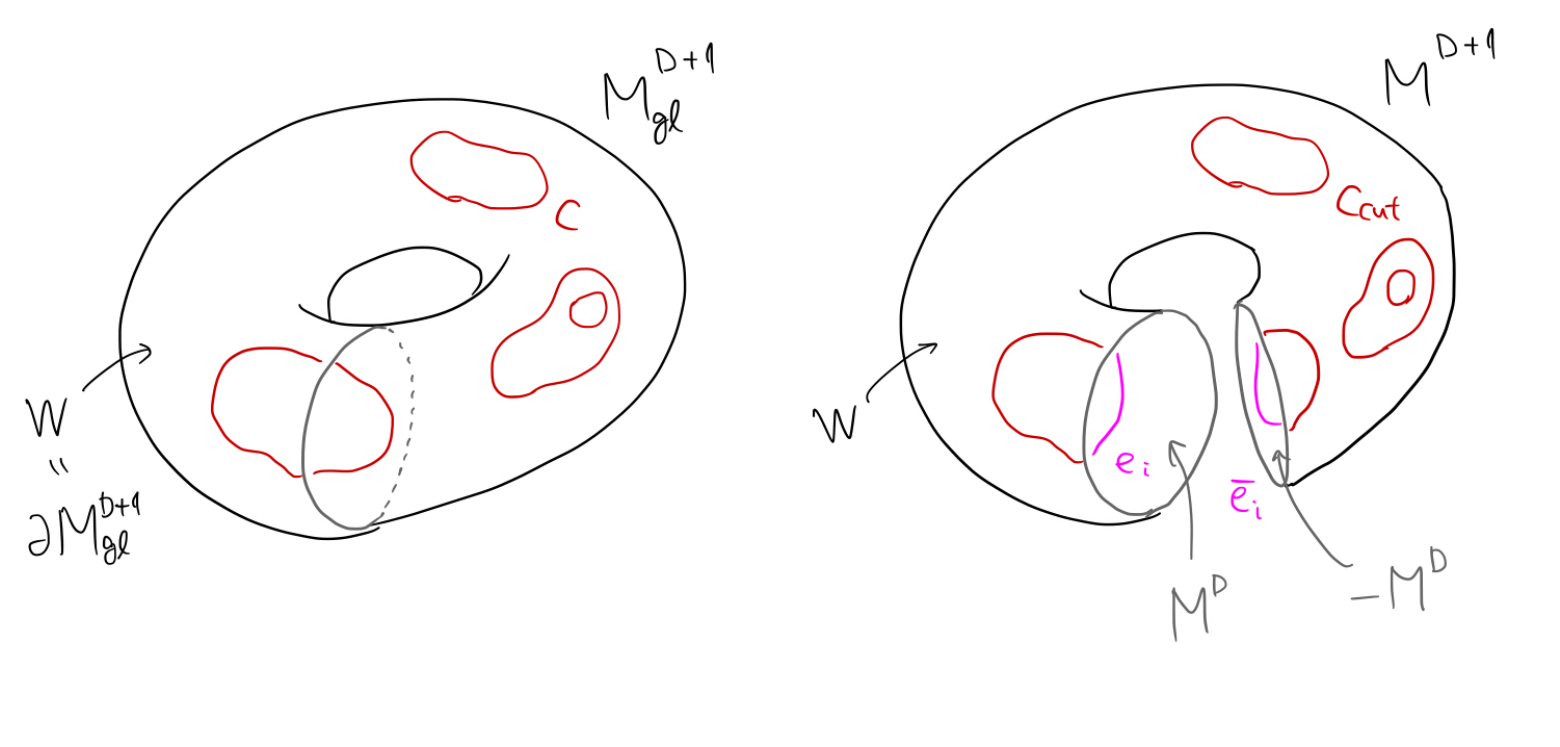}
\caption{Illustration of gluing relation.}
\label{fig:gluing}
\end{figure}

Then, we define the Hilbert space $\calV(M^{d}; c)$ as the configuration space modded out by equivalence relations (e.g., gauge transformations in discrete gauge theories),
\begin{align}
\calV(M^{d}; c):=\calC(M^{d}; c)/\sim.
\end{align}
The path integral $\calZ(M^{d+1})$ is a map from $\calV(\p M^{d+1})$ to a number,
\begin{align}
\calZ(M^{d+1}):\quad \calV(\p M^{d+1})\mapsto \mathbb{C}.
\end{align}
We will write this as $\calZ(M^{d+1})[c]$, for $c\in\calV(\p M^{d+1})$. 
The inner product in $\calV(M^{d}; c)$ is defined via bulk partition function as
\begin{align}
\langle x|y\rangle_{\calV(M^{d}; c)}:=\calZ(M^{d}\times I)[\overline{x}\cup y],
\end{align}
where $M^{d}\times I$ is a $(d+1)$-manifold pinched at $\partial M^d\times I$ by identification $(b, s)\sim(b, t)$ for $b\in\partial M^d$ and $s,t\in I$, 
so that $\partial(M^{d}\times I)=M\cup-M$. $\overline{x}$, $y$ specify boundary conditions on $-M$, $M$ respectively, where
$\overline{x}$ denotes the field configuration on $-M$ given by reversing orientation of $x$. Finally, we describe gluing relations for $(d+1)$-manifolds.
Let $M^{d+1}$ be a $(d+1)$-manifold whose boundary is $\p M^{d+1}=M^{d}\cup -M^{d}\cup W$, 
and $M^{d+1}_{\mathrm{gl}}$ be a $(d+1)$-manifold which is given by gluing the boundary of $M^{d+1}$ along $M^{d}$ and $-M^{d}$.
Then, the partition function $\calZ(M^{d+1}_{\mathrm{gl}})[c]$ on $M^{d+1}_{\mathrm{gl}}$ 
with the boundary condition $c\in\calV(W)$ on $W=\partial{M^{d+1}_{\mathrm{gl}}}$ is evaluated via the following gluing relation~\cite{barkeshli2016, Walker06},
\begin{align}
\calZ(M^{d+1}_{\mathrm{gl}})[c]=\sum_{e_i}\frac{\calZ(M^{d+1})[c_{\text{cut}}\cup e_i\cup \overline{e}_i]}{\langle e_i|e_i \rangle_{\calV(M^{d}; c_{\mathrm{cut}}^{d-1})}},
\label{gluing}
\end{align}
where $c_{\text{cut}}$ is the boundary condition inherited from $c$ after the cut, and $c_{\mathrm{cut}}^{d-1}$ is restriction of $c_{\text{cut}}$ to $\p M^{d}$.
$\{e_i\}$ is an orthonormal basis of $\calV(M^{d}; c_{\text{cut}}^{d-1})$.
We illustrate the gluing relation in Fig.\ref{fig:gluing}.

\subsection{$\calZ(L(k; 1,1,1))$: (3+1)D $C_k$ anomaly}
\label{subsec:Ck}
In this section, we explicitly compute $\calZ(L(k; 1,1,1))$, based on $(3+1)$D discrete gauge theory on the surface.
We pause here to mention that to construct (4+1)D SPT phases from given data of general (3+1)D surface theories.
We should first generalize the construction of (3+1)D bulk TQFT from (2+1)D topological ordered phases known as Walker-Wang construction~\cite{walker20123, von2013three},
to one dimension higher. The authors plan to do so in the future; in the present paper, we just exploit some machinery required to evaluate anomalies.

First, let us recall the definition of the lens space.
Let $k, p_j$ for $j=1,2,\dots n$ be natural numbers such that $\text{gcd}(k, p_j)=1$ for all $j$. 
The lens space $L(k; p_1,\dots, p_n)$ in $(2n-1)$D is defined as the quotient space by a free linear acton of cyclic group $\bZ_k$ on a sphere $S^{2n-1}$, 
considered as the unit sphere in $\mathbb{C}^n$. The $\bZ_k$ action is generated by
\begin{align}
(w_1, \dots, w_n)\mapsto(w_1\cdot e^{2\pi\im p_1/k}, \dots, w_n\cdot e^{2\pi\im p_n/k}).
\end{align}
Especially, $L(k;1,1,1)$ is a quotient space of $S^5$ by $\bZ_k$ action given by
\begin{align}
(w_1, w_2, w_3)\mapsto(w_1\cdot e^{2\pi\im /k}, w_2\cdot e^{2\pi\im /k}, w_3\cdot e^{2\pi\im /k}).
\end{align}

\subsubsection{handle decomposition}
\label{subsub:handle}

For evaluating the partition function, we perform handle decomposition of $L(k;1,1,1)$, which takes $L(k;1,1,1)$ apart into 5-balls. 
For $1\le k\le d$, $k$-handle in $d$ dimension is defined as a pair $(D^k\times D^{d-k}, S^{k-1}\times D^{d-k})$. 
$S^{k-1}\times D^{d-k}\subset\p{(D^k\times D^{d-k})}$ is called an attaching region of $k$-handle. 0-handle is defined as $D^d$.
We think of attaching $k$-handle to $d$-manifold $M_0$ with boundary, by an embedding of the attaching region 
$\phi: S^{k-1}\times D^{d-k}\mapsto\p M_0$ such that the image of $\phi$ is contained in $\p M_0$.
It is known that every compact $d$-manifold $M$ without boundary allows handle decomposition, i.e.,
$M$ is developed from a 0-handle by successively attaching to it handles of dimension $d$.

We find that the 5D lens space $L(k; 1,1,1)$ is decomposed into single $m$-handles for $m=0,1,2,3,4,5$.
We denote $L(k; 1,1,1)_m$ as the composition of $0, 1,2,\dots, m$ handles of $L(k; 1,1,1)$.

For convenience, we sometimes regard $L(k;1,1,1)$ as a $D^5$, 
whose points on the boundary $S^4=\p D^5$ are identified by a certain rule.
Concretely, we write $D^5$ as $D^4\times I$ ($D^4: \lvert w_1\rvert^2+\lvert w_2\rvert^2\le 1,\ $ $I: \theta_3\in[0,2\pi/k]$), 
pinched at $\p D^4\times I$. 
Then, $\p D^5$ consists of two $D^4$s at $\theta_3=0$ and $\theta_3=2\pi/k$, which is identified by the homeomorphism
\begin{align}
(w_1, w_2)\mapsto(w_1\cdot e^{2\pi\im /k}, w_2\cdot e^{2\pi\im /k}).
\label{D4identify}
\end{align}
We can obtain this picture, by seeing a $D^5$ as a subregion of $S^5$ such that $0\le\arg(w_3)\le 2\pi/k$.
If we define $\theta_3:=\arg(w_3)$, the $D^4$s at $\theta_3=0$ and $\theta_3=2\pi/k$ are indeed identified by~\eqref{D4identify}.
Now, the handle decomposition of $L(k; 1,1,1)$ is performed by the following steps:

\begin{enumerate}
\item First, we decompose $L(k; 1,1,1)$ into a 5-handle and $L(k; 1,1,1)_4$. 
We denote $\theta_1:=\text{arg}(w_1), \theta_2:=\text{arg}(w_2), \theta_3:=\text{arg}(w_3)$ for convenience.
The 5-handle is given by the subspace of $L(k;1,1,1)$ specified as
\begin{align}
\begin{split}
&\left\{(w_1, w_2, \theta_3) \middle| \lvert w_1\rvert\ge\epsilon, \quad \epsilon\le\theta_1\le\frac{2\pi}{k}-\epsilon\right\}, \\
&\left\{(w_1, w_2, \theta_3) \middle| \lvert w_1\rvert\ge\epsilon, \quad \frac{2\pi}{k}+\epsilon\le\theta_1\le\frac{4\pi}{k}-\epsilon\right\}, \\
&\dots \\
&\left\{(w_1, w_2, \theta_3) \middle| \lvert w_1\rvert\ge\epsilon, \quad \frac{2(k-1)\pi}{k}+\epsilon\le\theta_1\le2\pi-\epsilon\right\},
\end{split}
\end{align}
where $0<\epsilon\ll 1$ is a small positive constant. The above $k$ regions are connected by 
the identification map~\eqref{D4identify} with the neighboring one at $\theta_3=0$ and $\theta_3=2\pi/k$, making a single connected space.
This space is isomorphic to a subspace $0\le\theta_1\le 2\pi/k$ of the initial $S^5=\{(w_1, w_2, w_3)||w_1|^2+|w_2|^2+|w_3|^2=1\}$, thus $D^5$, as described before.

\item Next, we decompose $L(k; 1,1,1)_4$ into a 4-handle and $L(k; 1,1,1)_3$. The 4-handle is given by connected $k$ regions,
\begin{align}
\begin{split}
&\left\{(w_1, w_2, \theta_3) \middle| \lvert w_1\rvert\ge\epsilon, \quad -\epsilon\le\theta_1\le\epsilon\right\}, \\
&\left\{(w_1, w_2, \theta_3) \middle| \lvert w_1\rvert\ge\epsilon, \quad \frac{2\pi}{k}-\epsilon\le\theta_1\le\frac{2\pi}{k}+\epsilon\right\}, \\
&\dots \\
&\left\{(w_1, w_2, \theta_3) \middle| \lvert w_1\rvert\ge\epsilon, \quad \frac{2(k-1)\pi}{k}-\epsilon\le\theta_1\le\frac{2(k-1)\pi}{k}+\epsilon\right\}.
\end{split}
\end{align}
For fixed $\theta_1$, the above region looks like $\{\lvert w_1\rvert\ge\epsilon, \text{arg}(w_1)=0\}$ in the initial $S^5=\{(w_1, w_2, w_3)||w_1|^2+|w_2|^2+|w_3|^2=1\}$,
which specifies $D^4: \lvert w_2\rvert^2+ \lvert w_3\rvert^2\le 1-\epsilon^2$.
Hence, we see that the subspace makes a 4-handle $(D^4\times D^1, S^3\times D^1)$, where $D^1: \theta_1\in[-\epsilon, \epsilon]$. 

\item Let us examine what $L(k; 1,1,1)_3$ looks like. $L(k; 1,1,1)_3$ is a subspace of $L(k; 1,1,1)$ specified as $\lvert w_1\rvert\le\epsilon$.
As we did for $L(k;1,1,1)$, we can also parameterize $L(k; 1,1,1)_3$ by $(w_1, w_2, \theta_3)$ for $0\le \theta_3\le 2\pi/k$, embedded in $L(k;1,1,1)$. 
The subspace of $L(k; 1,1,1)_3$ given by fixing $\theta_3$ is a subregion $\lvert w_1\rvert\le\epsilon$ in $D^4: \lvert w_1\rvert^2+\lvert w_2\rvert^2\le 1$,
which is homeomorphic to $D^2\times D^2$. Since we are pinching at $\p D^4\times I$ in $L(k; 1,1,1)$, 
we also have to pinch at the subspace of $\p (D^2\times D^2)$ such that $\lvert w_1\rvert^2+\lvert w_2\rvert^2=1$. 
This pinching region becomes $D^2\times S^1\in\p (D^2\times D^2)$, where $S^1: \theta_2\in [0,2 \pi]$.
For convenience, we reparametrize $D^2\times D^2$ by $(z_1, z_2)$ such that $\lvert z_1\rvert\le 1, \lvert z_2\rvert\le 1$, 
where the pinching region is represented as $\lvert z_2\rvert= 1$.

Thus, $L(k; 1,1,1)_3$ is given by first preparing $(D^2\times D^2)\times I: (z_1, z_2, \theta_3)$, pinched at $(D^2\times S^1)\times I$.
Then, we identify $D^2\times D^2$s at $\theta_3=0$ and $\theta_3=2\pi/k$ by the map~\eqref{D4identify}, 
which is represented by $z_1, z_2$ as
\begin{align}
(z_1, z_2)\mapsto(z_1\cdot e^{2\pi\im /k}, z_2\cdot e^{2\pi\im /k}).
\label{D2D2identify}
\end{align}
If we forget about $z_1$, we can see that $(z_2, \theta_3)$ is essentially a parametrization of the 3D lens space $L(k; 1,1)$.
Namely, $L(k; 1,1)$ is quotient space of $S^3: \lvert z_2\rvert^2+\lvert z_3\rvert^2=1$ by the $\bZ_k$ action 
$(z_2, z_3)\mapsto(z_2\cdot e^{2\pi\im /k}, z_3\cdot e^{2\pi\im /k})$. 
If we pick up a subspace of $S^3$ specified by $0\le\theta_3:=\arg(z_3)\le2\pi/k$, 
we have $D^2\times I$ pinched at $S^1\times I$, with identification at $\theta_3=0$ and $\theta_3=2\pi/k$ 
by $z_2\mapsto z_2\cdot e^{2\pi\im /k}$, which corresponds to~\eqref{D2D2identify}. 
Hence, we can see $L(k; 1,1,1)_3$ as a fibre bundle whose base space is $L(k; 1,1)$, with fibre $D^2$.
Therefore, $L(k; 1,1,1)_3$ is the quotient space of $D^2\times S^3,  \{(z_1, (z_2, z_3)): \lvert z_1\rvert\le 1, \lvert z_2\rvert^2+\lvert z_3\rvert^2=1\}$ 
by the following $\bZ_k$ action denoted by $\sigma_k$,
\begin{align}
\sigma_k: (z_1, z_2, z_3)\mapsto(z_1\cdot e^{2\pi\im /k}, z_2\cdot e^{2\pi\im /k}, z_3\cdot e^{2\pi\im /k}).
\end{align}

The handle decomposition of $L(k; 1,1,1)_3$ is essentially given by decomposing the base space $L(k; 1,1)$ into handles, 
which is shown in Fig.\ref{fig:lenshandle}.
The 3-handle is given by the subspace of $(D^2\times D^2)\times I: (z_1, z_2, \theta_3)$ ($I: 0\le\theta_3\le 2\pi/k$), specified as
\begin{align}
\begin{split}
&\left\{(z_1, z_2, \theta_3) \middle| \lvert z_2\rvert\ge\epsilon, \quad \epsilon\le\arg(z_2)\le\frac{2\pi}{k}-\epsilon\right\}, \\
&\left\{(z_1, z_2, \theta_3) \middle| \lvert z_2\rvert\ge\epsilon, \quad \frac{2\pi}{k}+\epsilon\le\arg(z_2)\le\frac{4\pi}{k}-\epsilon\right\}, \\
&\dots \\
&\left\{(z_1, z_2, \theta_3) \middle| \lvert z_2\rvert\ge\epsilon, \quad \frac{2(k-1)\pi}{k}+\epsilon\le\arg(z_2)\le\frac{2(k-1)\pi}{k}-\epsilon\right\},
\end{split}
\end{align}
which are connected with each other by identification map~\eqref{D2D2identify}. 
This subspace gives a 3-handle $(D^3, S^2)$ in the base space $L(k; 1,1): (z_2, \theta_3)$.
Correspondingly, this gives a 3-handle $(D^3\times D^2, S^2\times D^2)$ in the total space $L(k; 1,1,1)_3=(D^2\times S^3)/\sigma_k$.

\begin{figure}[htb]
\centering
\includegraphics[bb=0 0 399 213]{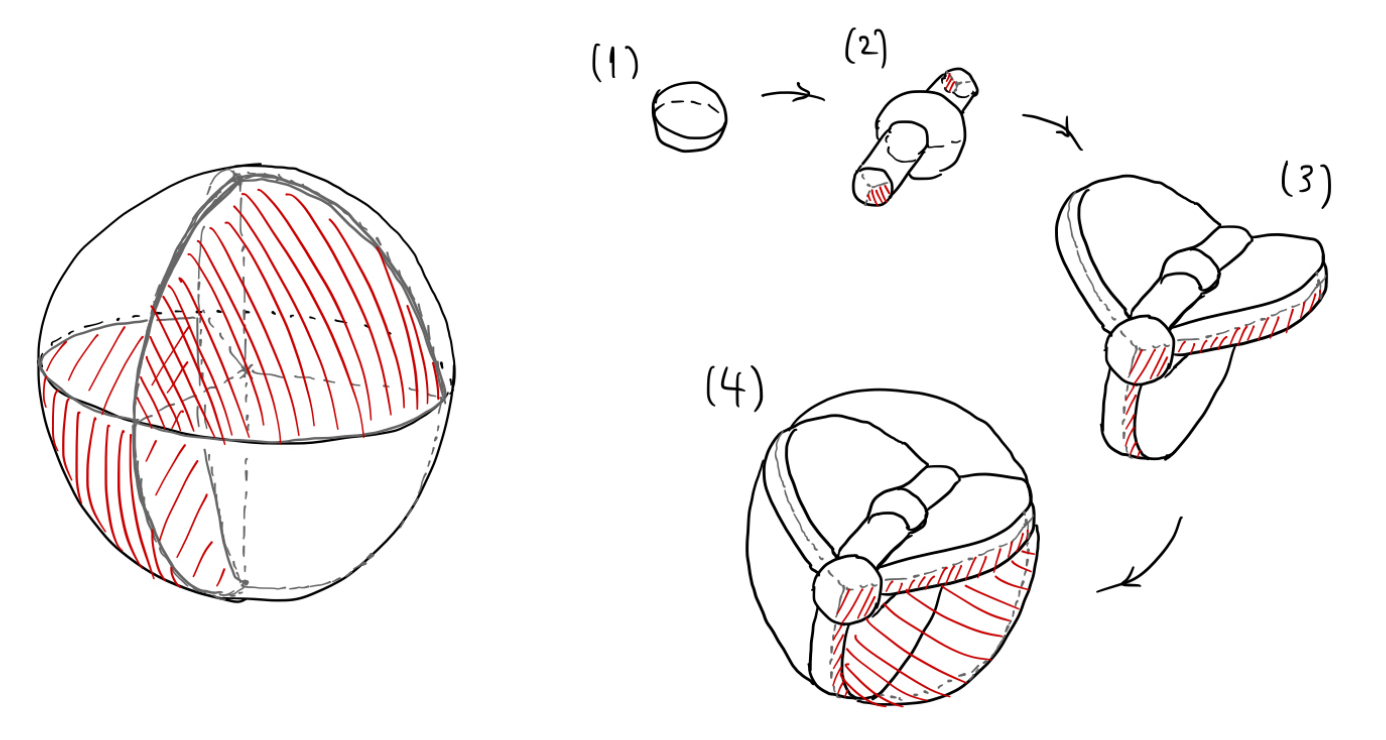}
\caption{3d Lens space $L(k;1,1)$ is obtained by identifying two $D^2$s on the boundary
 $\p D^3=D^2\cup D^2$, by $2\pi/k$ rotation $z_2\mapsto z_2\cdot e^{2\pi\im /k}$.
In the figure, we represent $L(k;1,1)$ for $k=3$, where the red regions are identified by this map. 
We can develop $L(3;1,1)$ by attaching handles successively.}
\label{fig:lenshandle}
\end{figure}

\item Then, we decompose $L(k;1,1,1)_2$ into a 2-handle and $L(k;1,1,1)_1$. The 2-handle is given by connected $k$ regions in $(D^2\times D^2)\times I$ parametrized by $(z_1, z_2, \theta_3)$
($I: 0\le\theta_3\le 2\pi/k$),
\begin{align}
\begin{split}
&\left\{(z_1, z_2, \theta_3) \middle| \lvert z_2\rvert\ge\epsilon, \quad -\epsilon\le\arg(z_2)\le\epsilon\right\}, \\
&\left\{(z_1, z_2, \theta_3) \middle| \lvert z_2\rvert\ge\epsilon, \quad \frac{2\pi}{k}-\epsilon\le\arg(z_2)\le\frac{2\pi}{k}+\epsilon\right\}, \\
&\dots \\
&\left\{(z_1, z_2, \theta_3) \middle| \lvert z_2\rvert\ge\epsilon, \quad \frac{2(k-1)\pi}{k}-\epsilon\le\arg(z_2)\le\frac{2(k-1)\pi}{k}+\epsilon\right\}.
\label{lens1handle}
\end{split}
\end{align}
For the base space $L(k; 1,1): (z_2, \theta_3)$ of $L(k; 1,1,1)_3=(D^2\times S^3)/\sigma_k$, 
the above subspace gives a 2-handle $(D^2\times D^1, S^1\times D^1)$.
Correspondingly, this gives a 2-handle $(D^2\times D^3, S^1\times D^3)$ in $L(k; 1,1,1)_3$.

Let us examine what $L(k; 1,1,1)_1$ looks like. $L(k; 1,1,1)_1$ is a subspace of $L(k; 1,1,1)_3=(D^2\times S^3)/\sigma_k$, 
specified as $\lvert z_2\rvert\le\epsilon$. In the base space $L(k; 1,1)$, $\{(z_2, \theta_3), \lvert z_2\rvert\le\epsilon\}$ gives 
a quotient space of $(D^2\times S^1), \{(z_2, \theta_3), \lvert z_2\rvert\le \epsilon, \theta_3\in\mathbb{R}/2\pi\bZ\}$ by the $\bZ_k$ action $\sigma_k$,
\begin{align}
\sigma_k: (z_2, \theta_3)\mapsto (z_2\cdot e^{2\pi\im/k}, \theta_3+2\pi/k).
\end{align}
Then, $L(k; 1,1,1)_1$ is a fibre bundle whose base space is given by $(D^2\times S^1)/\bZ_k$, with fibre $D^2$.
Therefore, $L(k; 1,1,1)_1$ is a quotient space of 
$(D^2\times D^2\times S^1), \{(z_1, z_2, \theta_3),  \lvert z_1\rvert\le1, \lvert z_2\rvert\le \epsilon, \theta_3\in\mathbb{R}/2\pi\bZ\}$ by the $\bZ_k$ action $\sigma_k$,
\begin{align}
\sigma_k: (z_1, z_2, \theta_3)\mapsto (z_1\cdot e^{2\pi\im/k}, z_2\cdot e^{2\pi\im/k}, \theta_3+2\pi/k).
\end{align}

\end{enumerate}

\subsubsection{computation of partition function}
Based on handle decomposition discussed above, now we evaluate $\calZ(L(k;1,1,1))$ via gluing relation.
In our case where the surface theory is described by discrete gauge theory, 
the boundary condition $\calC$ is an assignment of configuration of flat $G$-gauge field on boundaries.

\begin{enumerate}
\item First, we decompose $L(k;1,1,1)$ into a 5-handle and $L(k;1,1,1)_4$. 
The boundary condition on the attaching region $S^4$ is unique up to gauge equivalence,
since no surface or line operator can wrap $S^4$ nontrivially. 
Thus, the gluing relation becomes
\begin{align}
\begin{split}
\calZ(L(k;1,1,1))&=\frac{\calZ(L(k;1,1,1)_4)[\phi] \calZ(D^5)[\phi]}{\langle\phi|\phi\rangle_{\calV(S^4)}} \\
&=\frac{\calZ(L(k;1,1,1)_4)[\phi] \calZ(D^5)[\phi]}{\calZ(S^4\times D^1)[\phi]}, 
\label{L5L4}
\end{split}
\end{align}
where $\phi$ is the vacuum state on $S^4$. 
Similarly, for the decomposition of $L(k;1,1,1)_4$ into a 4-handle and $L(k;1,1,1)_3$, the gluing relation is expressed as
\begin{align}
\calZ(L(k;1,1,1)_4)[\phi]&=\frac{\calZ(L(k;1,1,1)_3)[\phi] \calZ(D^5)[\phi]}{\langle\phi|\phi\rangle_{\calV(S^3\times D^1)}}.
\label{L4L3}
\end{align}
Here, we have used again that no surface or line operator can wrap $S^3 \times D^1$ nontrivially. 
We can evaluate $\calZ(S^4\times D^1)[\phi]$ via gluing formula, by cutting $S^4\times D^1$ into two $D^5$s along $S^3\times D^1$,
\begin{align}
\calZ(S^4\times D^1)[\phi]=\frac{\calZ(D^5)[\phi]\calZ(D^5)[\phi]}{\langle\phi|\phi\rangle_{\calV(S^3\times D^1)}}.
\label{S4D1}
\end{align}
Combining~\eqref{L5L4}, \eqref{L4L3}, with~\eqref{S4D1}, we obtain
\begin{align}
\calZ(L(k;1,1,1))=\calZ(L(k;1,1,1)_3)[\phi].
\label{L5L3} 
\end{align}

\item Next, we decompose $L(k;1,1,1)_3$ into a 3-handle and $L(k;1,1,1)_2$. 
Now the attaching region is $S^2\times D^2$, where a surface operator can wrap $S^2$. 
Therefore, the boundary condition is labeled by a surface operator $W_{B, m}$ for $0\le m\le N-1$ wrapping $S^2$,
\begin{align}
\calZ(L(k;1,1,1)_3)[\phi]=\sum_{0\le m\le N-1}\frac{\calZ(L(k;1,1,1)_2)[W_{B, m}]\calZ(D^5)[W_{B, -m}]}{\langle W_{B, m}| W_{B, m}\rangle_{\calV(S^2\times D^2;\phi)}}. 
\label{L3L2}
\end{align}
Since a closed surface operator is a bubble on $\p D^5$, we have $\calZ(D^5)[W_{B, -m}]=\calZ(D^5)[\phi]$, since bubbles of surface operator for $\bZ_N$ gauge theory weights 1. 

$\langle W_{B, m}| W_{B, m}\rangle_{\calV(S^2\times D^2)}=\calZ(S^2\times D^3)[W_{B, -m}\cup W_{B, m}]$ is evaluated via gluing relation
by cutting $S^2\times D^3$ into two $D^5$s along $S^1\times D^3$, see Fig.\ref{fig:S2D3cut}. 
Here, the boundary condition on the cut is labeled by a line operator $W_{A, n}$ rounding $S^1$ of $S^1\times D^3$.
Hence, the gluing relation becomes
\begin{align}
\calZ(S^2\times D^3)[W_{B, -m}\cup W_{B, m}]=\sum_{0\le n\le N-1}\frac{\calZ(D^5)[W_{A, -n},W_{B, -m}]\calZ(D^5)[W_{A, n},W_{B, m}]}{\langle W_{A, n},e_{B, m}| W_{A, n},e_{B, m}\rangle_{\calV(S^1\times D^3;q_m, q_{-m})}}.
\label{eq:S2D3cut}
\end{align}
By the cutting, the surface operators before the cut $W_{B, -m}$, $W_{B, m}$ are
divided into two discs respectively. To make a membrane closed, the boundary condition $e_{B,m}$ on the cut $S^1\times D^3$ is 
introduced as a tube $S^1\times I$ connecting discs on $\p D^5$ (see Fig.\ref{fig:S2D3cut}).
Since bubbles of line and surface operator for $\bZ_N$ gauge theory weights 1, 
we have $\calZ(D^5)[W_{A, -n},W_{B, -m}]=\calZ(D^5)[\phi]$.
\footnote{When we have both line and surface operator on the boundary, one must care about linking of these two objects, 
since the correlator of these two objects has nontrivial phase~\eqref{BFbraid} when the line and surface are linked. 
In our case, we can have nontrivial linking in $\calZ(D^5)[W_{A, n},W_{B, m}]$ for some choice of configuration of operators. 
However, the linking in $\calZ(D^5)[W_{A, n},W_{B, m}]$ and $\calZ(D^5)[W_{A, n},W_{B, m}]$ cancels in the expression of~\eqref{eq:S2D3cut}, 
since the linking number is reversed for opposite orientation. In the main text, we are evaluating $\calZ(D^5)[W_{A, n},W_{B, m}]$ by choosing 
trivially linked configurations.} 
Moreover, by gluing relation, we can show that
\begin{align}
\langle W_{A, n},e_{B, m}| W_{A, n},e_{B, m}\rangle_{\calV(S^1\times D^3; q_m, q_{-m})}=1.
\label{norm1}
\end{align}
Therefore, we obtain
\begin{align}
\langle W_{B, m}| W_{B, m}\rangle_{\calV(S^2\times D^2; \phi)}=N\cdot\calZ(D^5)[\phi]\calZ(D^5)[\phi].
\end{align}
Combined with~\eqref{L3L2}, we have
\begin{align}
\calZ(L(k;1,1,1)_3)[\phi]=\frac{1}{N}\sum_{0\le m\le N-1}\frac{\calZ(L(k;1,1,1)_2)[W_{B, m}]}{\calZ(D^5)[\phi]}.
\label{L3L2final}
\end{align}

\begin{figure}[htb]
\centering
\includegraphics[bb=0 0 229 195]{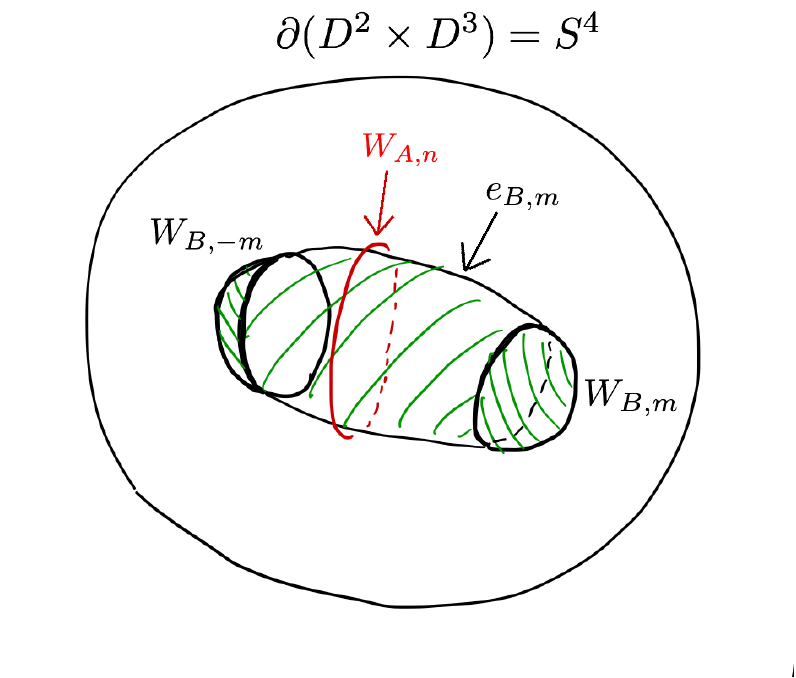}
\caption{The configuration of line and surface operator on $\p D^5$ after cutting $\calZ(S^2\times D^3)[W_{B, -m}\cup W_{B, m}]$ 
as~\eqref{eq:S2D3cut}. By the cutting, the surface operators before the cut $W_{B, -m}$, $W_{B, m}$ are
divided into two discs respectively. To make a membrane closed, the boundary condition $e_{B,m}$ on the cut $S^1\times D^3$ is 
introduced as a tube $S^1\times I$ connecting discs on $\p D^5$. Moreover, a line operator $W_{A,n}$ can round $S^1$ of the cut.}
\label{fig:S2D3cut}
\end{figure}

\item Then, we decompose $L(k;1,1,1)_2$ into a 2-handle and $L(k;1,1,1)_1$. 
Since the attaching region is $S^1\times D^3$, the boundary condition on the cut is labeled by a line operator rounding $S^1$.
The gluing relation becomes
\begin{align}
\calZ(L(k;1,1,1)_2)[W_{B, m}]=\sum_{0\le n\le N-1}\frac{\calZ(L(k;1,1,1)_1)[W_{A, n},W_{B, m}]\calZ(D^5)[W_{A, -n},W_{B, -m}]}{\langle W_{A, n},e_{B, m}| W_{A, n},e_{B, m}\rangle_{\calV(S^1\times D^3; q_m, q_{-m})}},
\end{align}
Using~\eqref{norm1}, we obtain
\begin{align}
\calZ(L(k;1,1,1)_2)[W_{B, m}]=\sum_{0\le n\le N-1}\calZ(L(k;1,1,1)_1)[W_{A, n},W_{B, m}]\calZ(D^5)[W_{A, -n},W_{B, -m}].
\label{L2L1final}
\end{align}
Combining~\eqref{L2L1final} with~\eqref{L3L2final}, we have
\begin{align}
\calZ(L(k;1,1,1)_3)[\phi]=\frac{1}{N}\sum_{0\le n, m\le N-1}\frac{\calZ(L(k;1,1,1)_1)[W_{A, n},W_{B, m}]\calZ(D^5)[W_{A, -n},W_{B, -m}]}{\calZ(D^5)[\phi]}.
\label{L3L1final}
\end{align}

\item
To evaluate $\calZ(L(k;1,1,1)_1)[W_{A, n},W_{B, m}]$, we should be careful about the configuration of line and surface operator, 
since the correlator of these two objects has nontrivial phase~\eqref{BFbraid} when the line and surface are linked.
To examine the configuration of these operators, we recall that $L(k; 1,1,1)_1$ is a quotient space of 
$(D^2\times D^2\times S^1), \{(z_1, z_2, \theta_3),  \lvert z_1\rvert\le1, \lvert z_2\rvert\le1, \theta_3\in\mathbb{R}/2\pi\bZ\}$ by the $\bZ_k$ action $\sigma_k$,
\begin{align}
\sigma_k: (z_1, z_2, \theta_3)\mapsto (z_1\cdot e^{2\pi\im/k}, z_2\cdot e^{2\pi\im/k}, \theta_3+2\pi/k).
\end{align}
We can choose the configuration of a surface operator $W_{B,m}$ as $(S^1\times S^1)/\sigma_k$ given by $\{z_1=0, \lvert z_2\rvert=1\}$.
Recall that the cut of $L(k;1,1,1)_2$ into $L(k;1,1,1)_1$ and a 2-handle is the subregion of $\p L(k; 1,1,1)_1$ given by
\begin{align}
\begin{split}
&\left\{(z_1, z_2, \theta_3) \middle| \lvert z_2\rvert= 1, \quad -\epsilon\le\arg(z_2)\le\epsilon\right\}.
\end{split}
\end{align}

Then, we know that a line operator on the cut $W_{A,n}$ can be located on $S^1$ given by $(f(\theta_3), 1, \theta_3)$ for $\theta_3\in\mathbb{R}/2\pi\bZ$, 
where $f:S^1\mapsto D^2$ is some function of $\theta_3\in\mathbb{R}/2\pi\bZ$.

Since we are putting a surface opeator $W_{B,m}$ at $z_1=0$, we must have $f\neq 0$ to make a line and surface operator dislocated 
(otherwise the linking number is ill-defined).
We choose $f$ as a constant function of $\theta_3$; $f(\theta_3)=p_0\neq0$, 
so that the linking of $W_{A,n}$, $W_{B,m}$ on the boundary of a 2-handle $\p D^5$ becomes trivial; $\calZ(D^5)[W_{A, -n},W_{B, -m}]=\calZ(D^5)[\phi]$
in~\eqref{L3L1final}.

To visualize the configuration of operators on $\p L(k;1,1,1)_1$, it is convenient to see $\p L(k;1,1,1)_1= (S^3\times S^1)/\sigma_k$ as a fibre bundle 
on a base space $S^1: 0\le\theta_3\le 2\pi/k$, with fibre $S^3$. At $\theta_3=2\pi/k$, we have a transition function on a fibre;
 $\sigma_k: (z_1,z_2)\mapsto(z_1\cdot e^{2\pi\im/k}, z_2\cdot e^{2\pi\im/k})$. 
If we regard $\theta_3$ as time direction, at a fixed time $\theta_3$ we have a loop-like excitation $q_m$ 
at $S^1:\{ z_1=0, \lvert z_2\rvert=1\}$ which corresponds to a time slice of $W_{B,m}$. 
We also have $k$ point-like excitations $e_n, C_k(e_n), \dots C_k^{k-1}(e_n)$ at 
$(z_1, z_2)=(p_0, 1), (p_0\cdot e^{2\pi\im/k}, e^{2\pi\im/k}), \dots,(p_0\cdot e^{2(k-1)\pi\im/k}, e^{2(k-1)\pi\im/k})$ respectively,
which correspond to time slice of $W_{A,n}$ (see Fig.\ref{fig:linking}). Remark that the label of excitation is transformed by $C_k$ associated with the transition function.
Especially, loop-like excitations are counted only when $C_k(q_m)=q_m$. 

We can see that $W_{A,n}$ and $W_{B,m}$ link exactly once; $\text{Lk}(W_{A,n}, W_{B,m})=1$, which is explained in Fig.\ref{fig:linking}.
\begin{figure}[htb]
\centering
\includegraphics[bb=0 0 530 184]{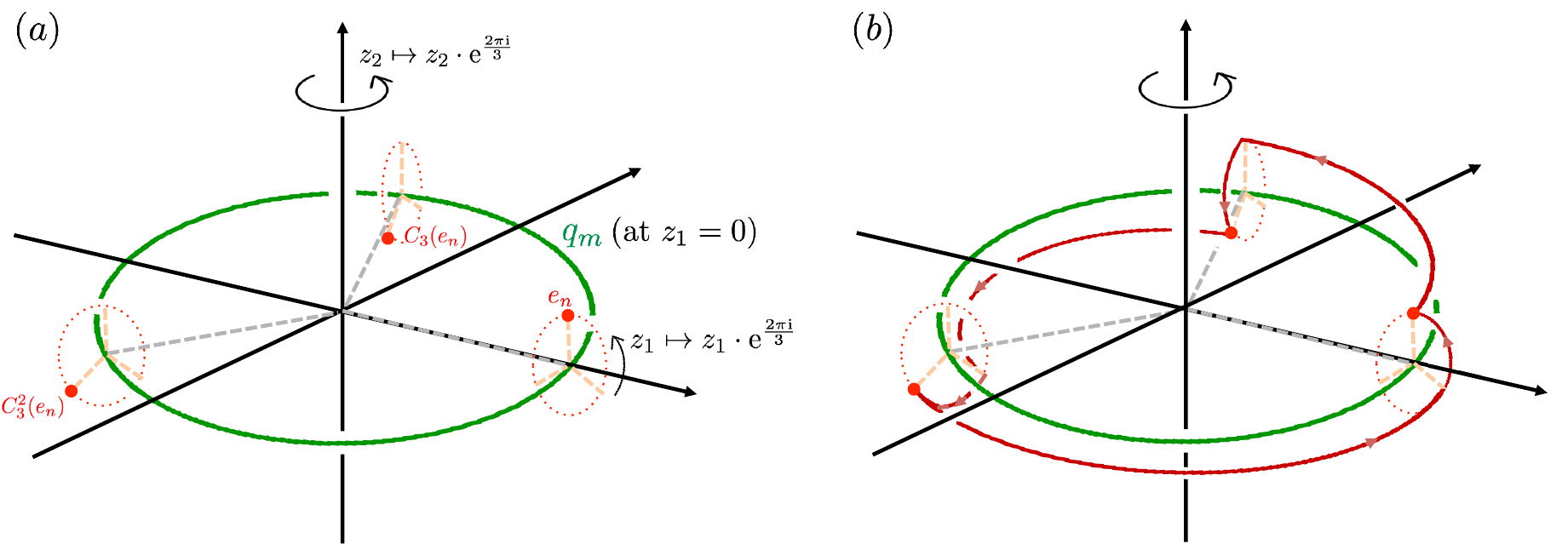}
\caption{$(a)$: The configuration of point-like and loop-like particles in $S^3$ is shown for $k=3$, 
via mapping to a unit 3-sphere $(z_1, z_2)\mapsto \left(z_1, z_2\sqrt{1-\lvert z_1\rvert^2}\right)$, and
stereograph mapping of unit 3-sphere $S^3\mapsto\mathbb{R}^3$.
In the stereograph projected picture, the action $z_2\mapsto z_2\cdot e^{2\pi\im/k}$ is realized as $2\pi/k$ rotation around $z$ axis in $\mathbb{R}^3$.
A loop-like excitation $q_m$ is located at $S^1$ given by $z_1=0$. In the vicinity of $z_1=0$, 
the action $z_1\mapsto z_1\cdot e^{2\pi\im/k}$ is realized as $2\pi/k$ rotation around $S^1$: $z_1=0$. For small $p_0$, 
the configuration of point-like excitations $e_n, C_k(e_n), \dots C_k^{k-1}(e_n)$ looks like red points, each of which is 
transformed to the neighboring one by acting $\sigma_k: (z_1,z_2)\mapsto(z_1\cdot e^{2\pi\im/k}, z_2\cdot e^{2\pi\im/k})$ associated with $C_k$ on particle label.
The configuration of these excitations are thus left invariant under the action of transition function $\sigma_k$, 
associated with $C_k$ on labels. 
$(b)$: To compute the linking number, it is convenient to think of the worldline
 of point-like particles (red line) transported gradually by the composite of $2\pi/k$ rotation, 
which finally amounts to the action of $\sigma_k$. As shown in the figure, the worldline links with a green loop of $q_m$ exactly once.
Since $\text{Lk}(W_{A,n}, W_{B,m})$ corresponds to the linking number of a time-independent $q_m$ loop and 
a world-line of $e_n$, we can see that $\text{Lk}(W_{A,n}, W_{B,m})=1$.
}
\label{fig:linking}
\end{figure}

Now, we have 
\begin{align}
\begin{split}
\calZ(L(k;1,1,1)_3)[\phi]&=\frac{1}{N}\sum_{0\le n, m\le N-1}\calZ(L(k;1,1,1)_1)[\{W_{A, n},W_{B, m}\}^{+1}] \\
&=\frac{1}{N}\sum_{0\le n, m\le N-1}\exp\left[-\frac{2\pi\im}{N}n m\right]\cdot\calZ(L(k;1,1,1)_1)[\{W_{A, n},W_{B, m}\}^{+0}],
\label{L3L1final_2}
\end{split}
\end{align}
where $+1$ means that $W_{A, n},W_{B, m}$ are linked once. 
The linking between these operators counts the extra factor of braiding phase $\exp\left[-\frac{2\pi\im}{N}n m\right]$~\eqref{BFbraid},
compared with the case where the line and surface operators are not linked, which is denoted as $\{W_{A, n},W_{B, m}\}^{+0}$.

Finally, let us consider gluing relation by cutting $L(k;1,1,1)_1=(D^4\times S^1)/\sigma_k$ into a $D^5$ along $D^4$ at $\theta_3=2\pi/k$.
The gluing relation becomes
\begin{align}
\begin{split}
&\calZ(L(k;1,1,1)_1)[\{W_{A, n},W_{B, m}\}^{+0}] \\
&=\eta_{e_n}\tilde{\eta}_{q_m}\cdot\sum_{e_i}\frac{\calZ(D^5)[\text{arc}(W_{A, n})\cup \text{arc}(W_{B, m})\cup e_i\cup \overline{e}_{i}]}{\langle e_i|e_i\rangle_{\calV(D^4)}},
\end{split}
\end{align}
where $\calV(D^4)$ is a shorthand notation of $\calV(D^4; q_m, e_n, C_k(e_n),\dots C_k^{k-1}(e_n))$,
which is a Hilbert space on the cut in the presence of excitations on its boundary $\p D^4$, as shown in Fig.\ref{fig:linking}.
$\{e_i\}$ is the orthonormal basis in $\calV(D^4)$ (it should not be confused with the notation of electric particle $e_n$).
$\text{arc}(W_{A,n}), \text{arc}(W_{B,n})$ denotes open line or surface operator after the cut.
Since we are acting $C_k$ on the cut, we count $C_k$ eigenvalues $\eta_{e_n}\tilde{\eta}_{q_m}$ on the Hilbert space of the cut $\calV(D^4)$.

$\calZ(D^5)[\text{arc}(W_{A, n})\cup \text{arc}(W_{B, m})\cup e_i\cup \overline{e}_{i}]$ contributes only when 
$k$ point-like particles fuses into vacuum and $C_k(q_m)=q_m$, otherwise weights zero.
Both $\calZ(D^5)[\text{arc}(W_{A, n})\cup \text{arc}(W_{B, m})\cup e_i\cup \overline{e}_{i}]$, $\langle e_i|e_i\rangle_{\calV(D^4)}$
reduces to evaluation of $\calZ(D^5)$ with unlinked bubbles of line and surface operators if 
$k$ point-like particles fuses into vacuum and $C_k(q_m)=q_m$, thus these factors become $\calZ(D^5)[\phi]$.
Therefore, we have
\begin{align}
\calZ(L(k;1,1,1)_1)[\{W_{A, n},W_{B, m}\}^{+0}]=\eta_{e_n}\tilde{\eta}_{q_m}\delta_{C_k^je_n}\delta_{q_m, C_k(q_m)},
\label{L1final}
\end{align}
where $\delta_{C_k^je_n}$ is 1 when $e_n, C_k(e_n),\dots C_k^{k-1}(e_n)$ fuses into vacuum, otherwize zero.

Combining~\eqref{L5L3}, \eqref{L3L1final_2} with~\eqref{L1final}, we finally obtain anomaly indicator
\begin{align}
\calZ(L(k;1,1,1))=\frac{1}{N}\sum_{n,m}\eta_{e_n}\tilde{\eta}_{q_m}\exp\left[-\frac{2\pi\im}{N}n m\right],
\label{lensabelian}
\end{align}
where the sum runs over $n$ such that $k$ particles $e_n, C_k(e_n), \dots C_k^{k-1}(e_n)$ fuse into vacuum, $m$ such that $q_m=C_k(q_m)$.

\end{enumerate}

\subsection{Example of anomalous $\bZ_N$ gauge theory}
Based on the indicator formula~\eqref{lensabelian}, we provide several examples of anomalous $\bZ_N$ gauge theories under $C_k$ symmetry.
First, let us think of the case where $C_k$ does not permute the label of quasiparticles. 
In this case, the electric particle $e_n$ contributes to the sum of the indicator~\eqref{lensabelian}, 
iff $k$ $e_n$ particles fuse into vacuum, i.e., $kn=0$ mod $N$. Thus, we can set $n$ as
\begin{align}
n=p\cdot\frac{N}{\gcd(k,N)},
\end{align}
where $p\in\bZ_{\gcd(k,N)}$. Let us define $\eta_p:=\eta_{e_{p\cdot{N}/{\gcd(k,N)}}}$. Since we have $(C_k)^k=1$, $\eta_p$ satisfies $(\eta_p)^k=1$.
In addition, $\eta$ should be compatible with fusion of quasiparticles: $\eta_a\eta_b=\eta_c$ if $a+b=c \mod \gcd(k,N)$. 
Therefore, $(\eta_p)^{\gcd(k,N)}=\eta_{e_0}=1$. To summarize, $\eta_p$ must satisfy
\begin{align}
(\eta_p)^{\gcd(k,N)}=1.
\end{align}

On the other hand, all vortex line operators $q_m$ contribute to the sum of~\eqref{lensabelian}, since $q_m=C_k(q_m)$ is always satisfied when $C_k$ does not permute labels.
Using the same logic as the case of $\eta_{p}$, we can see that $\tilde{\eta}$ satisfies 
$(\tilde{\eta}_{q_m})^k=1, (\tilde{\eta}_{q_m})^N=1$. Hence, $\tilde{\eta}_{q_m}$ must also satisfy
\begin{align}
(\tilde{\eta}_{q_m})^{\gcd(k,N)}=1.
\end{align}
Thus, we can set $\eta_p, \tilde{\eta}_{q_m}$ as
\begin{align}
\eta_p=\exp\left[\frac{2\pi\im}{\gcd(k,N)}\alpha p\right],\quad \tilde{\eta}_{q_m}=\exp\left[\frac{2\pi\im}{\gcd(k,N)}\beta m\right], 
\end{align}
where $\alpha,\beta\in\bZ_{\gcd(k,N)}$. Then, the indicator formula~\eqref{lensabelian} becomes
\begin{align}
\exp\left(\frac{2\pi\im\nu}{k}\right)=\frac{1}{N}\sum_{p,m}\exp\left[\frac{2\pi\im}{\gcd(k,N)}(\alpha p+\beta m-pm)\right]=\exp\left[\frac{2\pi\im}{\gcd(k,N)}\alpha\beta\right],
\label{nopermanomaly}
\end{align}
where the sum is taken over $p\in\bZ_{\gcd(k,N)}, m\in\bZ_N$. We can read the $\bZ_k$-valued anomaly from~\eqref{nopermanomaly} as
\begin{align}
\nu=\frac{k}{\gcd(k,N)}\alpha\beta \quad \mod k.
\end{align}

Next, we illustrate the case where labels of quasiparticles are changed by $C_k$ action. 
For instance, let us consider $\bZ_9$ gauge theory with $C_9$ symmetry $(N=k=9)$, and $C_k$ acts on quasiparticle labels as
\begin{align}
C_9: e_n\mapsto e_{4n}, \ q_m\mapsto q_{7m},
\end{align}
i.e., we have $r=4$, $s=7$ in~\eqref{ckaction1}.
We can check that the setup satisfies~\eqref{gcdcond},~\eqref{braidcond},~\eqref{ck1cond}. 
For this $C_9$ action, we can check that 9 particles $e_n, C_k(e_n),\dots C_9^{8}(e_n)$ fuse into vacuum for all $n\in\bZ_9$. 
Thus, all electric particles $e_n$ contribute to the sum of the indicator formula~\eqref{lensabelian}.
Since $\eta_{e_1}=\eta_{e_4}=(\eta_{e_1})^4$, $\eta_{e_1}$ satisfies $(\eta_{e_1})^3=1$.
Thus, we can express $\eta_{e_1}$ as $\eta_{e_1}=\omega^\alpha$, where $\omega=\e^{2\pi\im/3}$.

On the other hand, vortex lines $q_m$ contribute to the sum of~\eqref{lensabelian} iff $q_m=C_k(q_m)$. 
In our case, we can see that only $q_0, q_3, q_6$ are fixed under $C_k$.
Since $(\tilde{\eta}_{q_3})^3=1$, we can express $\tilde{\eta}_{q_3}$ as $\tilde{\eta}_{q_3}=\omega^\beta$.
Then, the indicator formula~\eqref{lensabelian} becomes
\begin{align}
\exp\left(\frac{2\pi\im\nu}{9}\right)=\frac{1}{9}\sum_{\substack{1\le n\le 9,\\[1pt]  1\le m\le 3}}\omega^{n\alpha+m\beta-nm}=\omega^{\alpha\beta}.
\end{align}
Therefore, the anomaly is read as $\nu=3\alpha\beta$ mod 9.

\section{Conjecture on non-abelian gauge theories}
\label{sec:nonabelian}
Finally, we pose a conjecture on $C_k$ anomaly indicator of (3+1)D gauge theories with non-abelian discrete gauge group. 
We consider (3+1)D untwisted Dijkgraaf-Witten theory, which supports both point-like and loop-like excitations~\cite{lan2018classification, delcamp2017};
\begin{itemize}
\item Point-like excitations

Point-like excitations are electric particles which can be created at the ends of an open line operator (Wilson line),
\begin{align}
W_{R_i}(C):=R_i\left(\prod_{ij\in C}g_{ij}\right),
\end{align}
where $C$ is an open line.
An electric charge is labeled by an irreducible representation $R_i\in\text{Rep}(G)$ of gauge group $G$, with quantum dimension $d_i=\dim[R_i]$.

\item Loop-like excitations

A single loop-like excitation can be created at the boundary of an open surface operator. 
In (3+1)D gauge theory, a vortex line excitation exists as a loop-like excitation.
A vortex line is characterized by a holonomy measured on a closed loop which rounds a vortex, 
which is labeled by a conjugacy class $\chi$ of $G$, since holonomy $h\in G$ is mapped $h\mapsto ghg^{-1}$ for some $g\in G$
under gauge transformation. A single vortex line is created by the following operator,
\begin{align}
M_{\chi}(S):=\sum_{h\in\chi}\left(\prod_{ij\in S}B_{ij}(h)\right),
\end{align}
where $S$ is an open surface on a dual lattice, and $B_{ij}$ is an operator which transforms a link variable $g_{ij}\mapsto hg_{ij}$. 
This operator implements a defect along $S$, and violates flatness at the boundary of $S$.
Quantum dimension can be defined as weight of a bubble of a surface operator $d_{\chi}:=\langle M_{\chi}(S)\rangle$, 
where $S$ is taken as a small sphere. Thus, we have $d_{\chi}=\lvert\chi\rvert$, where $\lvert\chi\rvert$ is the number of elements in $\chi$.

We can further think of attaching a charge on a vortex line labeled by $\chi$, defined as an irreducible representation $\text{Rep}_i(G_{\chi})$ of $G_{\chi}$,
where $G_{\chi}$ is a centralizer of $\chi$~\cite{lan2018classification, alford1992quantum, bucher1992topological}. 
Although we cannot explicitly write down the operator which creates such loop-like excitation associated with charge, 
we push on heuristic argument based on
the belief that charged loop-like excitation is generated by a composite object of Wilson line 
and surface operator. By including the effect of charge, the quantum dimension becomes
\begin{align}
d_{\chi; i}=\lvert\chi\rvert\cdot\dim[R_i(G_{\chi})].
\label{Gsurfaceop}
\end{align}

\end{itemize}
For example, let us consider $S_3$ gauge theory. $S_3$ is classified by three conjugacy classes; 
$\chi_1=((1)), \chi_2=((1,2),(2,3),(3,1)), \chi_3=((1,2,3), (1,3,2))$. For each conjugacy class, 
there are charged excitations labeled by irreducible representation of $G_{\chi_1}=S_3, G_{\chi_2}=\bZ_2, G_{\chi_3}=\bZ_3$ respectively.
$\chi_1$ corresponds to point-like excitations $p_1=(\chi_1, R_0(S_3)), p_2=(\chi_1, R_1(S_3)), p_3=(\chi_1, R_2(S_3))$ with quantum dimensions
$d_{1;0}=1, d_{1;1}=1, d_{1;2}=2$. $\chi_2, \chi_3$ corresponds to vortex lines, $s_{10}=(\chi_1, R_0(\bZ_2)), s_{11}=(\chi_1, R_1(\bZ_2)), 
s_{20}=(\chi_2, R_0(\bZ_3)), s_{21}=(\chi_2, R_1(\bZ_3)), s_{22}=(\chi_2, R_2(\bZ_3))$. Among them, 
$s_{10}=(\chi_1, R_0(\bZ_2))$ and $s_{20}=(\chi_2, R_0(\bZ_3))$ create ``pure'' vortex lines without charge, which are generated by~\eqref{Gsurfaceop}. 
The rest corresponds to bound states of electric charge and vortex line. 
Fusion rules of these excitations are controlled by fusion rule of quantum double $D(G)$~\cite{propitius1995topological, beigi2011quantum}
(i.e., inherit the fusion rules of anyons in (2+1)D $G$-gauge theory).

Based on heuristic computation of $\calZ(L(k;1,1,1))$ for given $G$-gauge theory on the surface, we conjecture that 
\begin{align}
\calZ(L(k;1,1,1))=\frac{1}{\lvert G\rvert}\sum_{\chi, i}d_{\chi; i}\eta_{\chi; i}\Theta_{\chi; i},
\label{nonabelindicator}
\end{align}
where $\eta_{\chi; i}$ is the $C_k$ eigenvalue of the state with a vortex line $\chi$ and 
$k$ charges $R_i(G_{\chi}), C_k[R_i(G_{\chi})], \dots, C_k^{k-1}[R_i(G_{\chi})]$ located in $C_k$ symmetric manner. 
$\Theta_{\chi;i}$ denotes the braiding phase between a charge and vortex line, $\Theta_{\chi;i}:=R_i(G_{h})[h]$, 
where $h\in\chi$ and $G_{h}$ is the centralizer of $h$. ($\Theta_{\chi;i}$ becomes a scalar due to Schur's lemma.)
The sum in~\eqref{nonabelindicator} runs over vortex lines $\chi$ fixed by $C_k$ action $\chi=C_k[\chi]$, and electric charges such that $k$ particles 
$R_i(G_{\chi}), C_k[R_i(G_{\chi})], \dots, C_k^{k-1}[R_i(G_{\chi})]$ fuse into vacuum (i.e., tensor product of $k$ representations contains the trivial representation of $G_{\chi}$). The detail of the heuristic argument is found in Appendix~\ref{app:nonind}.

\section{conclusion and outlook}
In summary, we have proposed anomaly indicator for (3+1)D discrete gauge theories enriched with $C_k$ rotation symmetry with proof.
The symmetry fractionalization properties of point and loop-like excitations are summarized as $\eta, \tilde{\eta}$, and
the indicator formula enables us to evaluate anomalies immediately from given data of symmetry fractionalization.
Some future generalizations would be to examine the generalization for anomalies based on other point group symmetries.
Another possible generalization is to look for the indicator formula of anomalies in (3+1)D that correspond to $\Omega_5^{SO}(pt)=\bZ_2$.
It would also be interesting to see the generalization for fermionic topological phases.

\acknowledgments

R.K. thank Yuji Tachikawa and Chang-Tse Hsieh for useful discussions. 
R.K. is supported by Advanced Leading Graduate Course for Photon Science (ALPS) of Japan Society for the Promotion of Science (JSPS).
K.S. is supported by PRESTO, JST (JPMJPR18L4).

\appendix
\section{$C_k$ anomaly of discrete non-abelian gauge theories}
\label{app:nonind}
In this Appendix, we heuristically compute the path integral of (4+1)D SPT phases, with discrete non-abelian gauge theory on the surface.
We would like to evaluate partition function $\calZ(L(k;1,1,1))$ as we did for $\bZ_N$ gauge theory in the main text.
As exercise, let us first try to compute partition function on a sphere, $\calZ(S^5)$.
\subsection{$\calZ(S^5)$}
First, we apply gluing relation by cutting $S^5$ along $S^4$ into two $D^5$s,
\begin{align}
\calZ(S^5)=\frac{\calZ(D^5)[\phi]\calZ(D^5)[\phi]}{\calZ(S^4\times D^1)[\phi]}.
\end{align}
Similarly, by cutting $S^4\times D^1$ along $S^3\times D^1$ into two $D^5$, we have
\begin{align}
\calZ(S^4\times D^1)=\frac{\calZ(D^5)[\phi]\calZ(D^5)[\phi]}{\calZ(S^3\times D^2)[\phi]}.
\end{align}
Combining these relations gives
\begin{align}
\calZ(S^5)=\calZ(S^3\times D^2)[\phi]. 
\label{S5S3D2}
\end{align}
$S^3\times D^2$ can be decomposed into two $D^5$s by cutting along $S^2\times D^2$. 
Now, the boundary condition on the cut $S^2\times D^2$ is labeled by a surface operator wrapping $S^2$.
Since no line operator can round $S^2\times D^2$ nontrivially, these surface operators cannot carry electric charge. 
Thus, the boundary condition is characterized by ``pure'' surface operators without charge, $s_{\chi;0}$ wrapping $S^2$.
Therefore, by the gluing relation
\begin{align}
\calZ(S^3\times D^2)[\phi]=\sum_{\chi}\frac{\calZ(D^5)[s_{\chi;0}]\calZ(D^5)[s_{\chi;0}]}{\langle s_{\chi;0}|s_{\chi;0}\rangle_{\calV(S^2\times D^2;\phi)}}.
\label{GS3D2}
\end{align}
Since the configuration of $s_{\chi;0}$ of $\calZ(D^5)[s_{\chi;0}]$ is a bubble on $\p D^5$,
it follows that $\calZ(D^5)[s_{\chi;0}]=\lvert\chi\rvert\calZ(D^5)[\phi]$. For $\langle s_{\chi;0}|s_{\chi;0}\rangle_{\calV(S^2\times D^2;\phi)}$, we have
\begin{align}
\langle s_{\chi;0}|s_{\chi;0}\rangle_{\calV(S^2\times D^2;\phi)}=\calZ(S^2\times D^3)[s_{\chi;0}\cup s_{\overline{\chi};0}].
\end{align}
We have two pure surface operators on the boundary. $\calZ(S^2\times D^3)$ is evaluated by cutting along $S^1\times D^3$. 
Now the boundary condition on the cut contains a line operator rounding $S^1$, 
together with a tube $e_{\chi;0}$ of surface operator $s_{\chi;0}$ supported on $S^1\times I$ (see Fig.\ref{fig:S2D3cut}).
Such configuration of surface operators attached to a line operator are labeled by $(\chi, R_i(G_{\chi}))$.
Thus, the gluing relation becomes
\begin{align}
\calZ(S^2\times D^3)[s_{\chi;0}\cup s_{\overline{\chi};0}]=\sum_{R_i\in\text{Rep}(G_{\chi})}\frac{\calZ(D^5)[s_{\chi;i}]\calZ(D^5)[s_{\chi;i}]}{\langle e_{\overline{\chi};\overline{i}}|e_{\chi;i}\rangle_{\calV(S^1\times D^3; (\chi; R_i)\cup\overline{(\chi, R_i)})}},
\end{align}
where $e_{\chi;i}$ denotes a tube of surface operator attached to a line operator with irreducible representation $R_i(G_{\chi})$.
We can show by gluing relation that $\langle e_{\overline{\chi};\overline{i}}|e_{\chi;i}\rangle_{\calV(S^1\times D^3; (\chi; R_i)\cup\overline{(\chi, R_i)})}=1$, hence
\begin{align}
\begin{split}
\calZ(S^2\times D^3)[s_{\chi;0}\cup s_{\overline{\chi};0}]&=\sum_{R_i\in\text{Rep}(G_{\chi})} \left(\lvert\chi\rvert\cdot\dim[R_i(G_{\chi})]\right)^2\cdot\calZ(D^5)[\phi]\calZ(D^5)[\phi] \\
&=\lvert G_{\chi}\rvert\cdot(\lvert\chi\rvert\calZ(D^5)[\phi])^2. 
\label{GS2D3}
\end{split}
\end{align}
Combining~\eqref{GS2D3} with~\eqref{GS3D2}, we obtain
\begin{align}
\begin{split}
\calZ(S^5)=\sum_{\chi}\frac{1}{\lvert G_{\chi}\rvert}=\sum_{\chi}\frac{\lvert\chi\rvert}{\lvert G\rvert}=1.
\end{split}
\end{align}
Thus, we have $\calZ(S^5)=1$~\cite{yonekura2018cobordism}, which is required for cobordism invariance of (4+1)D path integral.~\cite{freedhopkins}

\subsection{$\calZ(L(k;1,1,1))$}
Next, we evaluate partition function on the 5D lens space. Using the logic of~\ref{subsec:Ck}, we obtain~\eqref{L5L3},
\begin{align}
\calZ(L(k;1,1,1))=\calZ(L(k;1,1,1)_3)[\phi].
\end{align}
Next, we do handle decomposition for $L(k;1,1,1)_3$,
\begin{align}
\begin{split}
\calZ(L(k;1,1,1)_3)[\phi]&=\sum_{\chi}\frac{\calZ(L(k;1,1,1)_2)[s_{\chi;0}]\calZ(D^5)[s_{\chi;0}]}{\langle s_{\chi;0}|s_{\chi;0}\rangle_{\calV(S^2\times D^2;\phi)}} \\
&=\sum_{\chi}\frac{\calZ(L(k;1,1,1)_2)[s_{\chi;0}]}{\lvert G_{\chi}\rvert\cdot\lvert\chi\rvert\calZ(D^5)[\phi]} \\
&=\frac{1}{\lvert G\rvert}\sum_{\chi}\frac{\calZ(L(k;1,1,1)_2)[s_{\chi;0}]}{\calZ(D^5)[\phi]},
\label{GL3L2}
\end{split}
\end{align}
where we used~\eqref{GS2D3}.
For $L(k;1,1,1)_2$,
\begin{align}
\begin{split}
\calZ(L(k;1,1,1)_2)[s_{\chi;0}]&=\sum_{R_i\in\text{Rep}_i(G_{\chi})}\frac{\calZ(L(k;1,1,1)_1)[s_{\chi;i}^{(+1)}]\calZ(D^5)[s_{\chi;i}]}{\langle e_{\overline{\chi};\overline{i}}|e_{\chi;i}\rangle_{\calV(S^1\times D^3; (\chi; R_i)\cup\overline{(\chi, R_i)})}}\\
&=\sum_{R_i\in\text{Rep}_i(G_{\chi})}\calZ(L(k;1,1,1)_1)[s_{\chi;i}^{(+1)}]\cdot d_{\chi;i}\cdot\calZ(D^5)[\phi],
\label{GL2L1}
\end{split}
\end{align}
where we choose the configuration of surface and line operator of $s_{\chi;i}$ in $\calZ(L(k;1,1,1)_1)[s_{\chi;i}^{(+1)}]$, such that these operators link 
exactly once on the surface of $L(k;1,1,1)_1$,
as explained in the main text (see the step 4 of Section \ref{subsec:Ck}, and Fig.\ref{fig:linking}).

Therefore, using the logic in the final step of Section \ref{subsec:Ck}, we have
\begin{align}
\calZ(L(k;1,1,1)_1)[s_{\chi;i}^{(+1)}]=\sum_{\chi,i}\Theta_{\chi;i}\eta_{\chi;i},
\end{align}
where $\eta_{\chi;i}$ is an eigenvalue of $C_k$ for the state with a single vortex line $\chi$ and $k$ charges $R_i$ located in rotation symmetric manner.
$\Theta$ denotes a phase by linking between a line and a surface; $\Theta_{\chi;i}:=R_i(h)$, where $h\in\chi$ and we define $G_\chi$ as a centralizer of $h$. 
Then, $R_i(h)$ becomes a scalar because of Schur's lemma~\cite{propitius1995topological}. The sum runs over $\chi$ such that $C_k(\chi)=\chi$ for a vortex line, and $R_i$ such that 
$R_i, C_k(R_i), \dots, C_k^{k-1}(R_i)$ can fuse into vacuum. Finally, we obtain
\begin{align}
\calZ(L(k;1,1,1))=\frac{1}{\lvert G\rvert}\sum_{\chi,i}d_{\chi;i}\Theta_{\chi;i}\eta_{\chi;i}.
\end{align}



\bibliography{ref}

\end{document}